\documentclass[12pt]{article}

\usepackage{newtxtext,newtxmath}

\usepackage{amsmath,amssymb,amsfonts}%
\usepackage{bm}
\usepackage{braket}
\usepackage{multirow, booktabs,multicol}
\usepackage{caption, subcaption}
\usepackage[normalem]{ulem}
\usepackage{graphicx}

\usepackage[letterpaper,margin=1in]{geometry}

\renewenvironment{abstract}
	{\quotation}
	{\endquotation}

\date{}

\makeatletter
\renewcommand{\fnum@figure}{\textbf{Figure \thefigure}}
\renewcommand{\fnum@table}{\textbf{Table \thetable}}
\makeatother

\usepackage{scicite}

\usepackage{url}

\newcommand{\dd}{\mathrm{d}} 

\DeclareMathOperator*{\argmin}{arg\,min}

\newcommand{\mn}{\mathbf{n}}
\newcommand{\mbE}{\mathbb{E}}

\usepackage{CJKutf8}

\def\scititle{
Solving the Hubbard model with Neural Quantum States
}
\title{\bfseries \boldmath \scititle}

\author{
	Yuntian Gu$^{1,2\dagger}$,
	Wenrui Li$^{1,2\dagger}$,
        Heng Lin$^{2,3\dagger}$,
	Bo Zhan$^{2,4\dagger}$,\and
        Ruichen Li$^{1,2}$,
        Yifei Huang$^{2}$,
        Di He$^{1\ast}$, %
        Yantao Wu$^{4\ast}$,\and %
        Tao Xiang$^{4\ast}$, %
        Mingpu Qin$^{5\ast}$, %
        Liwei Wang$^{1\ast}$, %
        Dingshun Lv$^{2\ast}$ %
    \and
	\small$^{1}$State Key Laboratory of General Artificial Intelligence, School of \and
    \small Intelligence Science and Technology, Peking University. \\
	\small$^{2}$ByteDance Seed, China.\and
        \small$^{3}$State Key Laboratory of Low-Dimensional Quantum Physics, Department of Physics, Tsinghua University.\and
	\small$^{4}$Institute of Physics, Chinese Academy of Sciences.\and
        \small$^{5}$Key Laboratory of Artificial Structures and Quantum Control (Ministry of Education), \and
        \small School of Physics and Astronomy, Shanghai Jiao Tong University, Shanghai, China.\and
	\small$^\ast$Corresponding author: dihe@pku.edu.cn (D. He);  yantaow@iphy.ac.cn (Y. T. Wu);\and \small txiang@iphy.ac.cn (T. Xiang); qinmingpu@sjtu.edu.cn(M. P. Qin); \and \small wanglw@pku.edu.cn (L. W. Wang); lvdingshun@bytedance.com (D. S. Lv).
	\and \small$^\dagger$These authors contributed equally to this work.
}

\begin{document} 

\maketitle

\begin{abstract} \bfseries \boldmath

The rapid development of neural quantum states (NQS) has established it as a promising framework for studying quantum many-body systems. In this work, by leveraging the cutting-edge transformer-based architectures and developing highly efficient optimization algorithms, we achieve the state-of-the-art results for the doped two-dimensional (2D) Hubbard model, arguably the minimum model for high-Tc superconductivity. Interestingly, we find different attention heads in the NQS ansatz can directly encode correlations at different scales, making it capable of capturing long-range correlations and entanglements in strongly correlated systems. With these advances, we establish the half-filled stripe in the ground state of 2D Hubbard model with the next nearest neighboring hopping, consistent with experimental observations in cuprates. Our work establishes NQS as a powerful tool for solving challenging many-fermions systems.

\end{abstract}

\noindent

The Hubbard model \cite{J.Hubbard, qin2022hubbard,arovas2022hubbard} is the iconic model for studying the correlation effect in many-electron systems. Though its Hamiltonian is simple, the physics it can harbor is extremely rich, including quantum magnetism, Mott insulator, charge and spin density waves, and so on. Following the discovery of cuprate high-Tc superconductors, the two-dimensional (2D) Hubbard model has attracted increasing attention due to its potential connection to the microscopic mechanism of high-Tc superconductivity \cite{doi:10.1126/science.235.4793.1196,PhysRevB.37.3759}. The study of the 2D Hubbard model relies mainly on numerical many-body methods due to the lack of analytic solutions. In the past few decades, a variety of many-body methods were developed, including density matrix renormalization group (DMRG)~\cite{PhysRevLett.69.2863} and its higher dimension generalization in the language of tensor network states \cite{RevModPhys.93.045003,xiang2023density}, such as projected entangled pair states (PEPS)~\cite{verstraete2004}, quantum embedding methods such as dynamical mean-field theory~(DMFT)\cite{RevModPhys.68.13}, density matrix embedding theory~(DMET)\cite{PhysRevLett.109.186404}, and different flavors of quantum Monte Carlo approaches \cite{Gubernatis_Kawashima_Werner_2016,Becca_Sorella_2017,AFQMC-lecture-notes-2013} and so on. In the past, different methods usually gave inconsistent results due to the limitations of each method \cite{PhysRevX.5.041041}. In recent years, large scale cross-validation has been performed among different methods, and consensus has begun to reach about the ground state of the doped 2D Hubbard model, with \cite{science.aam7127} and without the next nearest neighboring hopping \cite{xu2024coexistence}.

\begin{figure*}[]
\centering
\includegraphics[width=\textwidth]{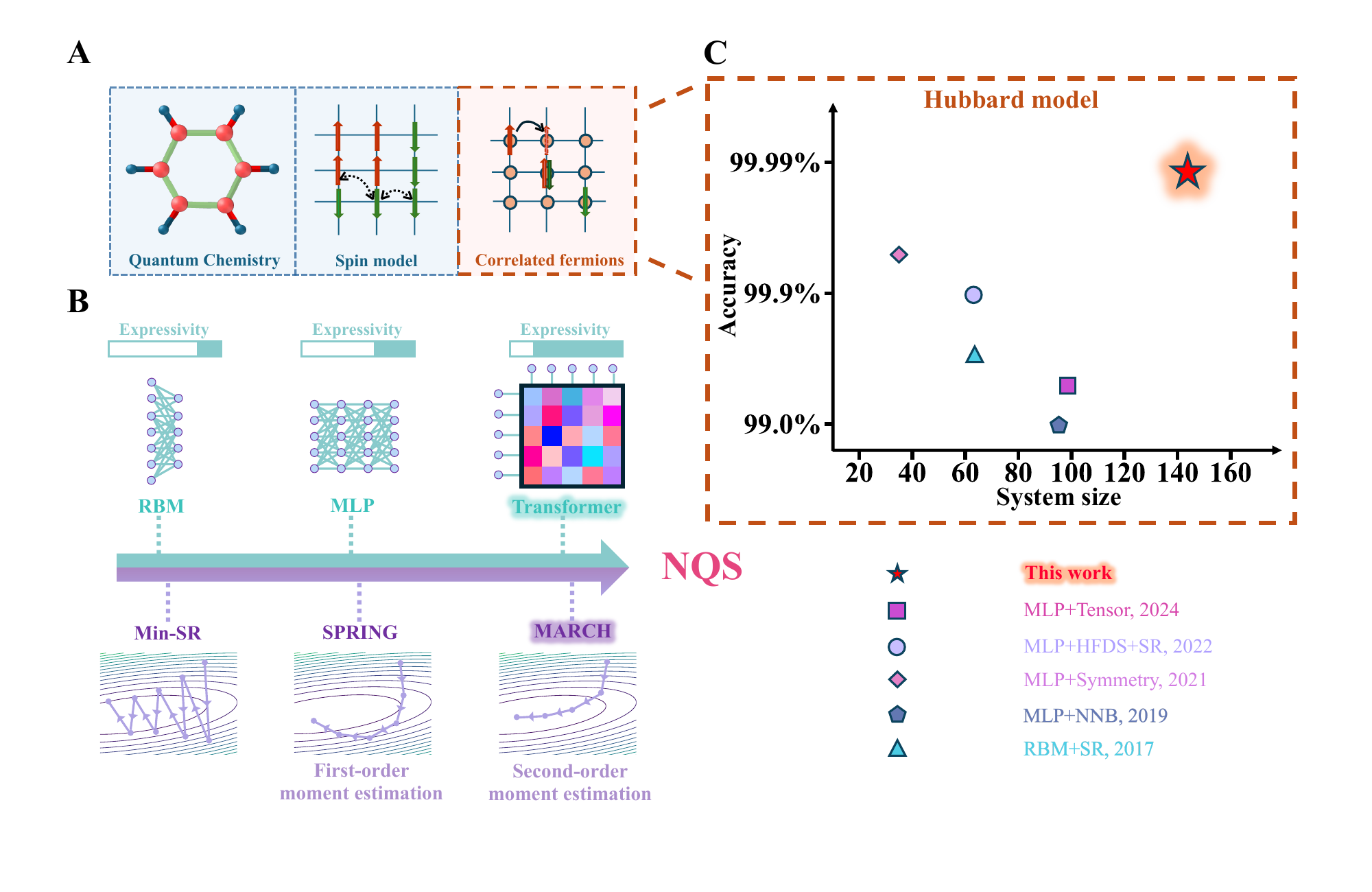}
\caption{\textbf{Application, architecture, optimization, and performance of Neural Quantum States (NQS) on the Hubbard model}. (A) Guided by the variational principle, NQS are powerful and versatile, applicable to a wide range of quantum systems, including ab initio quantum chemistry, spin models, and correlated fermionic models. (B) Various neural network architectures can be used to construct NQS, with increasing expressivity from Restricted Boltzmann Machines (RBM) \cite{nomura2017restricted}, Multi-Layer Perceptrons (MLP) \cite{inui2021determinant, luo2019backflow, robledo2022fermionic, zhou2024solving} to Transformers. These networks are optimized using methods inspired by quantum information, including Min-SR~\cite{chen2024empowering, rende2024simple}, SPRING~\cite{goldshlager2024kaczmarz}, and the advanced MARCH method proposed in this work. (C) A performance comparison of different NQS methods on the 2D Hubbard model, plotting simulation accuracy against system size. We benchmark the accuracy on systems where exact solutions are available (see Supplementary Section~\ref{sec:acc_estimate} for details). This work achieves higher accuracy on larger system sizes than previously reported in other notable NQS studies, demonstrating a significant advancement in the field.
}
\label{fig1}
\end{figure*}

On another front, researchers are also striving to develop more accurate and efficient new methods, aiming to provide more reliable results for strongly correlated systems like the Hubbard model to unveil the underlying exotic physics. In recent years, benefiting from the rapid development in the fields of neural networks and machine learning, Neural Quantum States (NQS) have emerged as a promising new approach for solving quantum many-body systems.
 Significant progress has been made in the development of advanced NQS methods in quantum systems since their inception by Carleo et al.~\cite{carleo2017solving}, which %
have markedly improved the accuracy of NQS approaches, establishing them as a powerful framework for investigating systems in quantum chemistry \cite{pfau2020ab, ren2023towards, li2024computational} and interacting spin models \cite{nomura2021dirac, roth2023high, yu2024solving}, as shown in Figure~\ref{fig1}(A). In light of these significant achievements, a crucial question arises: how effective is this new paradigm for the study of the more challenging strongly correlated fermionic lattice models?

In this direction, extensive effort has been made to investigate the utilization of NQS methods in the context of the Hubbard model, particularly in the most challenging 2D lightly doped cases.
 Early attempts to apply NQS approaches to the 2D Hubbard model\cite{nomura2017restricted} faced challenges in accurately representing fermionic nodal structures. Subsequent work by Luo et al.\cite{luo2019backflow} introduced a neural network backflow ansatz that substantially improved the representation power of the wavefunction, demonstrating its effectiveness on systems as large as $12\times 8$.
More recently in 2022, Robledo et al.~\cite{robledo2022fermionic} achieved highly accurate results for various geometries of the Hubbard model up to $8\times 8$ by combining Hidden-Fermion Determinant States (HFDS) with Stochastic Reconfiguration (SR) \cite{sorella2001generalized, nightingale2001optimization, sorella2007weak}.
Despite this notable progress, scaling NQS simulations to even larger, more physically relevant parameter regimes is hindered by network architectural limitations and optimization barriers\cite{zhou2024solving, karthik2024convolutional}.

In this work, we make significant progress in expanding the scalability and accuracy of the NQS approach for solving the Hubbard model with two key innovations as shown in Figure~\ref{fig1}(B). First, a novel NQS architecture is proposed, which leverages the self-attention mechanism~\cite{vaswani2017attention}, a technique proven to be very effective in scaling up~\cite{kaplan2020scaling}. The built-in long-range structure makes the NQS ansatz different from tensor network states, which consist of only local tensors.  This ensures that the NQS can precisely capture long-range correlations and entanglements in strongly correlated systems and can accurately handle larger systems, which is infeasible for other neural network architectures like Restricted Boltzmann Machine (RBM) \cite{nomura2017restricted} and Multi-Layer Perceptron (MLP) \cite{inui2021determinant, luo2019backflow, robledo2022fermionic, zhou2024solving}.  Second, we propose an enhanced optimization algorithm, termed Moment-Adaptive ReConfiguration Heuristic (MARCH), designed to achieve substantially faster and more stable convergence than previous optimization algorithms such as SR~\cite{sorella2001generalized, nightingale2001optimization, sorella2007weak}, Min-SR \cite{chen2024empowering, rende2024simple} and SPRING \cite{goldshlager2024kaczmarz}, especially for large and frustrated systems. 

With these architectural and algorithmic advancements, we achieve state-of-the-art variational ground state energy of the 2D Hubbard model at unprecedented sizes up to $16\times 16$, for both open and periodic boundary conditions, as shown in Figure~\ref{fig1}(C). As a result, we confirm the filled stripe ground state in the pure Hubbard model and identify a half-filled stripe ground state with the next nearest neighboring hopping $t'=-0.2$, consistent with experimental observations in cuprate~\cite{tranquada1995evidence}. Moreover, in the $t'=-0.2$ case, we find the stripe is horizontal, i.e., 
stripe arrangements prefer the longer side of the geometry. This is quite different from previous calculations on cylinders \cite{science.aam7127}, indicating the effectiveness of our method in eliminating the boundary effects. We also have a careful analysis of the structure of our NQS ansatz and find that different
attention heads can directly encode correlations at different scales after optimization, making it capable to capture long-range correlations and entanglements in the Hubbard model. Our developments in architectures and optimization strategies establish NQS as a powerful and transformative paradigm for solving the Hubbard model. This framework extends readily to a broader range of strongly correlated systems, offering a new lens to explore new exotic quantum many-body states resulting from strong correlations.

\subsection*{The Hubbard model}

The Hamiltonian of the Hubbard model is:

\begin{equation}\label{Hamil}
\hat H = -t\sum _{\langle ij \rangle, \sigma} \hat c _{i\sigma} ^\dagger \hat c _{j\sigma} - t ^ \prime \sum _{\langle \langle ij \rangle \rangle, \sigma} \hat c _{i\sigma} ^\dagger \hat c _{j\sigma} + U \sum _i \hat n _{i \uparrow} \hat n _{i \downarrow}.
\end{equation}
Here $\hat c _{i\sigma} ^\dagger$ $(\hat c _{j\sigma})$ is the creation~(annihilation) operator for an electron with spin $\sigma$ at site $i(j)$, $\hat n_{i\sigma} = \hat c _{i\sigma} ^\dagger \hat c _{i\sigma} $ is the particle number operator. The terms $\langle ij \rangle$ and $\langle \langle ij \rangle \rangle$ denote pairs of nearest neighboring and next nearest neighboring sites, respectively. We study the 2D Hubbard model defined on square lattices with size $L_x \times L_y$. Both open boundary conditions (OBC) and periodic boundary conditions (PBC) are adopted in this work. We set the nearest neighboring hopping amplitude $t=1$ as the unit of energy. The on-site Coulomb repulsion is fixed at $U=8$, a value relevant to cuprates~\cite{andersen1995lda, hirayama2018ab}. We consider both the pure Hubbard case ($t^\prime=0$) and the case with the next nearest neighboring hopping ($t^\prime=-0.2$). In this work, we focus on the lightly doped regime with hole concentration of $\delta =1/8$, which is the most intriguing region in both the 2D Hubbard model and in cuprates.

\subsection*{Methods and benchmarks}

\begin{figure*}[]
\centering
\includegraphics[width=0.95\textwidth]{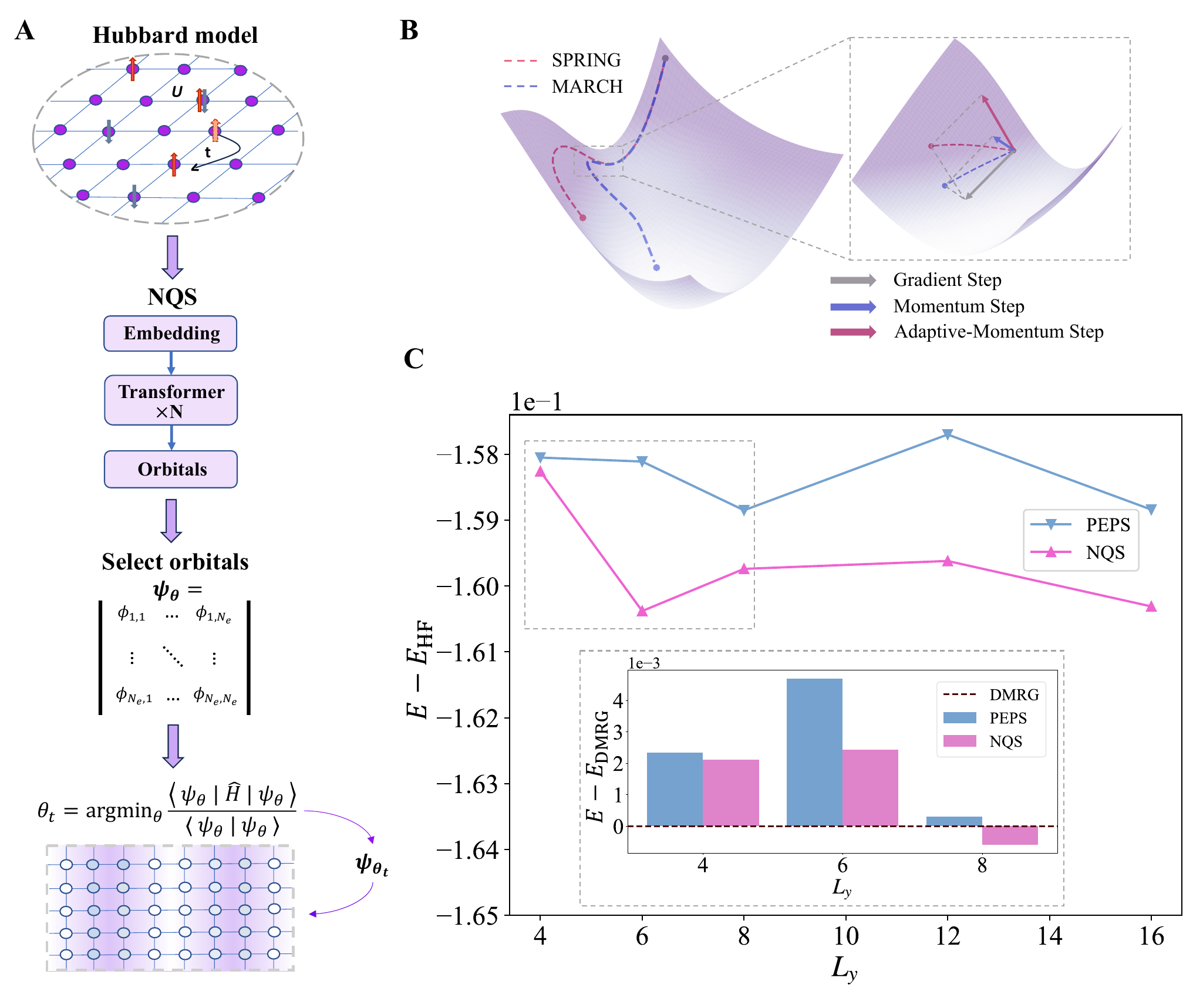}
\caption{\textbf{NQS architecture, optimization scheme, and benchmark performance on the Hubbard model.}
(\textbf{A}) An input lattice configuration is processed through a transformer-based neural network to generate backflow orbitals. These orbitals are then used to construct the NQS wavefunction $\psi _\theta$. The variational parameters $\theta$ are optimized to find the ground state.
(\textbf{B}) Optimization trajectories for MARCH and SPRING. MARCH follows a more direct path by dynamically adapting its step size for each parameter. The inset shows how SPRING's reliance on momentum might lead to overshooting, whereas MARCH's adaptive-momentum step prevents oscillations and accelerating convergence to the minimum.
(\textbf{C}) Comparison of ground-state energies with PEPS (bond dimension $D \geq 20) $ \cite{liu2025} for the pure Hubbard model on $16\times L_y$ system with OBC. Our NQS results consistently achieve lower variational energies than PEPS  across different lattice geometries. The inset shows we also outperform reference DMRG energies \cite{liu2025} (with $32000$ SU(2) multiplets) when $L_y = 8$.
}
\label{fig2}
\end{figure*}

Figure~\ref{fig2} presents an overview of our NQS algorithm. As shown in Figure~\ref{fig2}(A), we employ a transformer-based NQS, parameterized by $\theta$, to represent the amplitude $\psi_{\theta}(\mn)$ associated with each configuration $\ket{\mn}$. Our transformer-based NQS assigns a unique high-dimensional embedding vector to each site $i$. These embeddings are subsequently processed through a series of multi-head attention blocks, the core component of modern large language models~\cite{brown2020language}. The attention block is defined as:
\begin{align}
\label{eq:attn}
    \text{Attn}(Q,K,V)=\text{softmax}(QK^{\top} / \sqrt{d_H})V,
\end{align}
where $Q, K, V\in\mathbb{R}^{N\times d_H}$ are feature matrices produced by neural networks. Here, $N$ represents the number of sites in the Hubbard model, $d_H$ is the hidden dimension of the attention block. The $i$-th row of these matrices corresponds to the $i$-th site in the Hubbard model, arranged in dictionary order. The normalized attention scores $S_{\text{attn}}$, given by $\text{softmax}(QK^{\top} / \sqrt{d_H})$, allow the block to aggregate information from all sites by assigning weights. 
Thus, the attention block can capture complex, long-range correlations, which is crucial for modeling strongly correlated systems like the Hubbard model. The multi-head attention block employs distinct feature matrices to capture diverse correlation patterns in the system. After self-attention layers, a linear layer is employed to produce backflow orbitals $M(\mn)$. The amplitude is then derived as the determinant of orbitals occupied by electrons.

In our work, we aim to find the ground state by minimizing the total energy of the wavefunction. In previous NQS studies \cite{carleo2017solving}, SR-based optimizers have been used to minimize the energy. These optimizers effectively simulate the imaginary time evolution, enhancing the convergence rate over the standard first-order optimizer by exploiting the quantum geometric tensor~\cite{cheng2010quantum}. One particular method is the SPRING algorithm \cite{goldshlager2024kaczmarz}, which leverages the history of gradients to accelerate convergence and smooth out the optimization path. Building on this, we introduce the MARCH algorithm. MARCH enhances SPRING by also incorporating an estimate of the second moment of the gradients, drawing a direct parallel to how the popular Adam optimizer~\cite{kingma2017adammethodstochasticoptimization} improves upon Momentum~\cite{sutskever2013importance}. This addition of second-moment information is crucial for navigating the complex and high-dimensional energy landscapes that typically exist in quantum systems. It provides an estimate of the gradient's variance, allowing MARCH to adapt the learning rate for each parameter individually. This adaptive capability offers a significant advantage, particularly when encountering challenging regions like saddle points or flat plateaus where the gradient is small. While a momentum-based method like SPRING might overshoot, MARCH can use the second-moment information to maintain progress and ``march" confidently out of these areas. As illustrated in Figure~\ref{fig2}(B), this results in a more direct and efficient path to the energy minimum, leading to faster and more robust convergence.

These combined architectural and algorithmic advancements enable us to accurately tackle 2D Hubbard model with unprecedented sizes beyond the reach of previously NQS methods. We first perform benchmark calculations for systems under PBC for the pure Hubbard model ($t^\prime = 0$) at half-filling, where numerical exact ground state energies can be obtained by Auxiliary Field Quantum Monte Carlo (AFQMC)~\cite{qin2016benchmark}. The NQS energies are consistent with the AFQMC results within twice the statistical errors (see Table~\ref{tab:afqmc} in the Supplementary Materials).

We then benchmark the new NQS method on the doped pure Hubbard model under OBC with varying sizes $L_x$ and $L_y$. Direct comparisons of ground-state energies with state-of-the-art DMRG and PEPS~\cite{liu2025} results are presented in Figure~\ref{fig2}(C) (with Hartree-Fock energies as reference). In narrow systems such as $L_y \leq 6$, the DMRG energy is the lowest. However, when reaching the 2D limit by increasing the width of the system, where DMRG fails to capture the area law of entanglement entropy, the NQS results surpass DMRG. Noticeably, in the entire range of $L_y$ we studied, NQS achieves lower variational energies than those reported by state-of-the-art PEPS with bond dimension $D \geq 20$ \cite{liu2025}.

 \begin{figure*}
     \centering
     \includegraphics[width=1\linewidth]{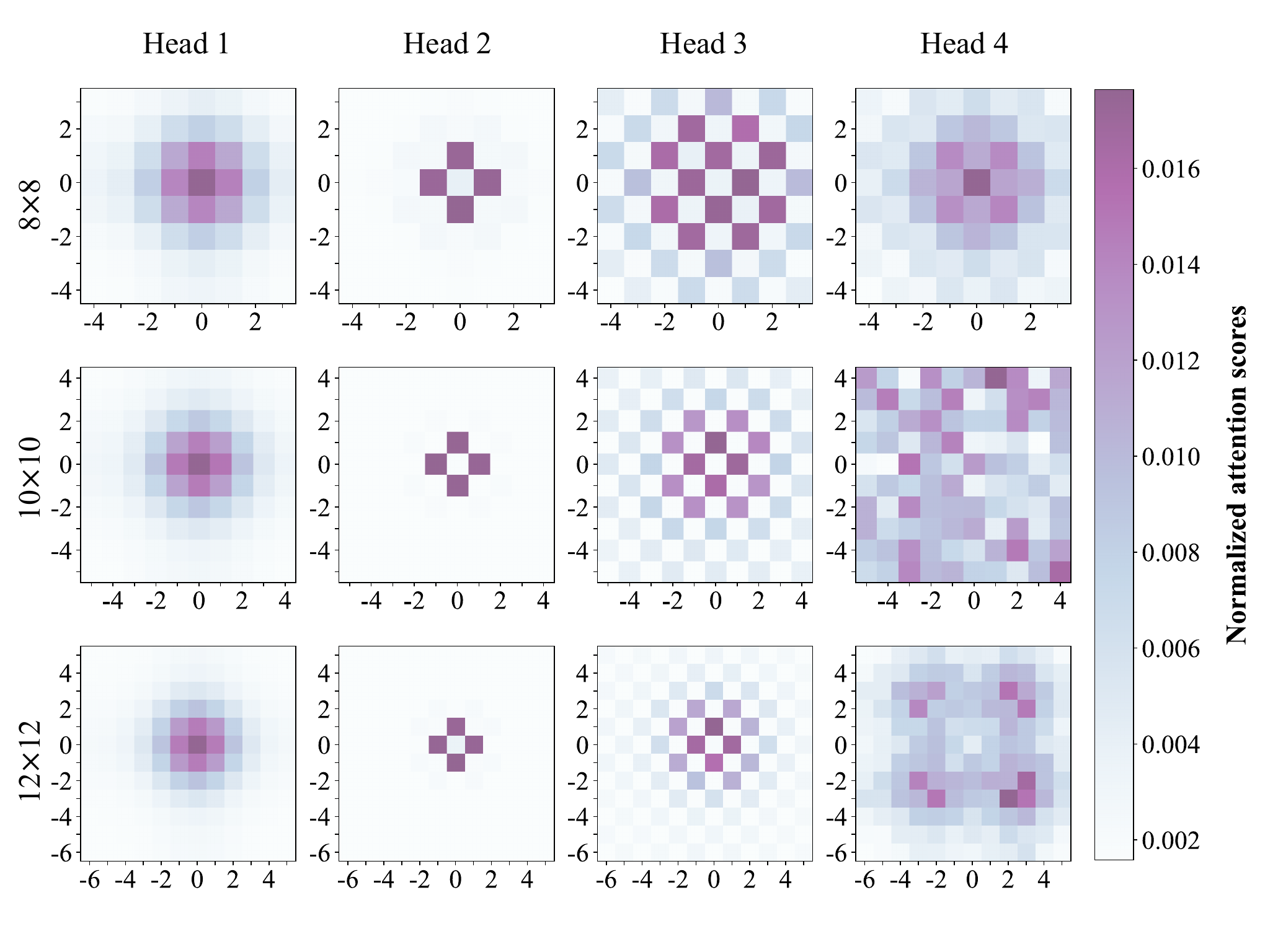}
     \caption{\textbf{Visualization of the attention mechanism.} Shown is how a lattice site attends to another target site in the 2D pure Hubbard model for lattice sizes of $8\times 8$, $10 \times 10$, and $12 \times 12$ at half filling. The four attention heads, which are equivalent and have been rearranged by pattern for clarity, each correspond to a specific physical aspect: short-range correlations, nearest-neighbor hopping, antiferromagnetic correlations, and complex and delocalized correlations in the ground state.}
     \label{fig4}
 \end{figure*}

To gain deeper insight into how the transformer architecture captures the intricate physics of the Hubbard model, we examine the attention patterns learned by the network \cite{viteritti2025transformer}. In our transformer ansatz, the heads after optimization are found to specialize in capturing the correlations in different scales present in the ground-state wave function. This emergent specialization is analogous to observations in fields like natural language processing, where different attention heads in a transformer learn distinct, interpretable patterns such as syntactical dependencies~\cite{vaswani2017attention}. In Figure \ref{fig4}, we visualize the normalized attention scores $S_{\text{attn}}$ for the half-filling systems, which quantify the learned pairwise importance between sites, from the first layer for representative systems. Similar results for the doped cases or deeper layers can be found in Figures~\ref{fig:score 8x8 half}-\ref{fig:score 12x12 doped site} in the Supplementary Materials. The results are striking: the first head clearly learns to prioritize local interactions, with $S_{\text{attn}}$ that decay with the distance between sites, capturing short-range correlations. The second head specializes in nearest-neighbor hopping, with its attention focusing almost exclusively on the four adjacent sites, which is a fundamental process in the Hubbard model. The third head successfully identifies the emergent antiferromagnetic order, assigning high attention scores in a checkerboard pattern characteristic of this magnetic correlation. The final head appears to capture more complex, delocalized phase information, exhibiting a sophisticated pattern that is crucial for accurately representing the full complexity of the wave function. This specialization demonstrates the remarkable ability of the transformer's attention mechanism to autonomously discover and disentangle the fundamental physical correlations governing the system.

\subsection*{Filled stripe order in pure Hubbard model}

 \begin{figure*}
     \centering
     \includegraphics[width=1\linewidth]{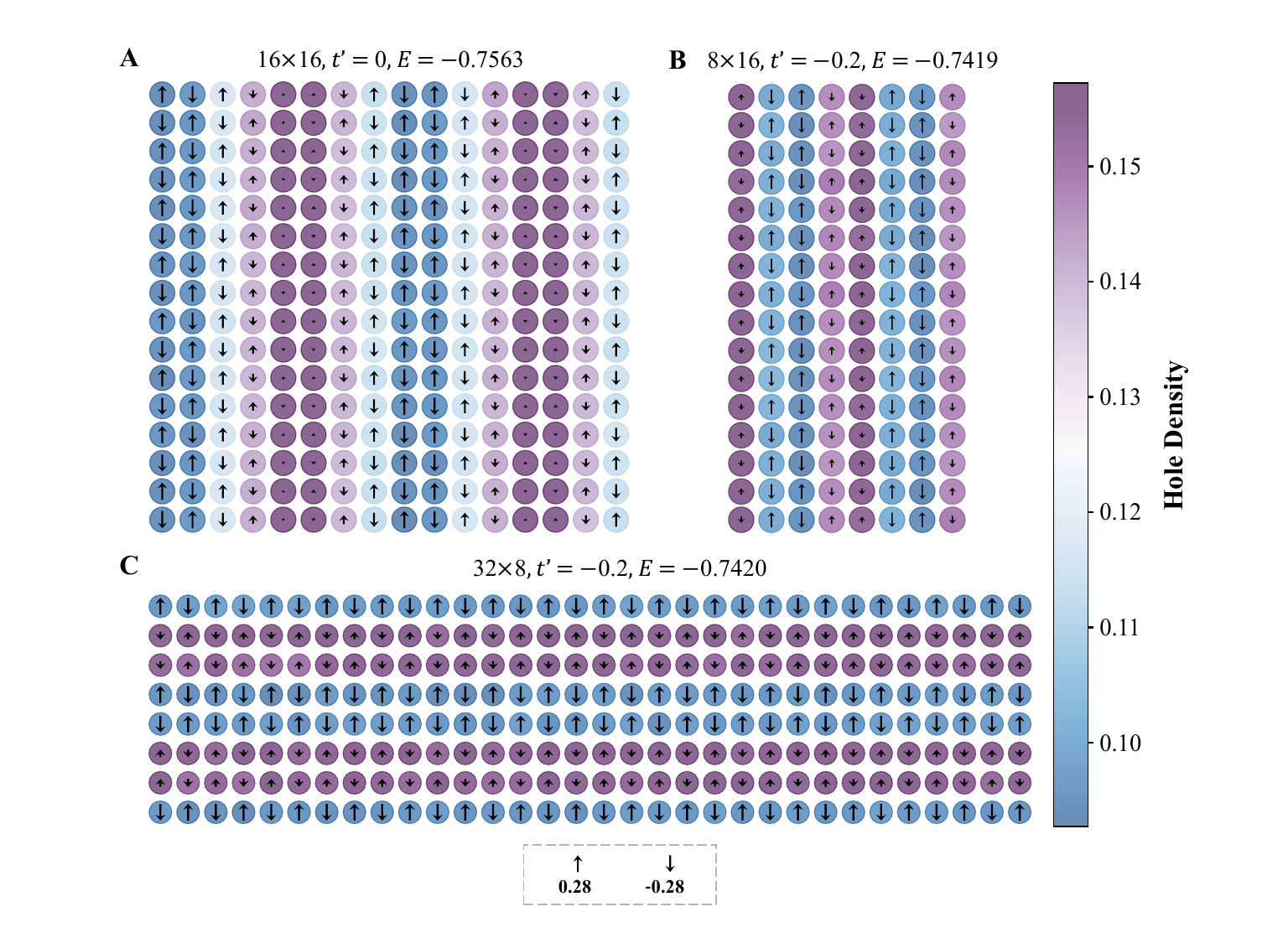}
     \caption{\textbf{Hole and spin density distributions in the ground state for different lattice sizes with $U = 8$ and hole doping $\delta = 1/8$ under PBC in the Hubbard model}. The magnitudes of the spin density are represented by the sizes of the arrows while the direction is denoted by the direction of the arrows. Hole density is depicted using a color scale. Stripes can be clearly seen in all three systems. (A) Results for $16 \times 16$ pure Hubbard model where the stripe state with $\lambda=8$ is observed. (B) Results for $16 \times 12$ system with $t'=-0.2$ where the stripe state with $\lambda=4$ is observed. (C) Results for $32 \times 8$ system with $t'=-0.2$ where the stripe state with $\lambda=4$ is observed. Notice that the stripes in the ground state for $t^\prime = -0.2$ in (B) and (C) are horizontal, even though both vertical and horizontal stripes are allowed in the systems. More results can be found in the supplementary materials.}
     \label{fig:PBC 16x16, PBC t2 16x12, 32x8 Combined}
 \end{figure*}

The ground state of the lightly doped pure ($t^\prime = 0$) 2D Hubbard model was established as the filled stripe before \cite{science.aam7127}. Recently Liu et al.~\cite{liu2025} observed a crossover of the stripe direction with the increase of system widths for systems under OBC, indicating the existence of a large finite size effect in systems with open boundaries. We performed systematic calculations of the pure Hubbard model for different sizes under both OBC and PBC. For systems with OBC, we find stripe states that match well with the results obtained from PEPS but with lower energies (see Figure~\ref{fig2}(C)). In particular, our results also demonstrate a crossover between horizontal and vertical stripes as a function of system width (from $L_y=8$ to $L_y=12$, see Figures~\ref{fig:OBC 16x8}-\ref{fig:OBC 16x12} in the Supplementary Materials). But when we switch to PBC, the crossover vanishes (see Figures~\ref{fig:PBC 16x8}-\ref{fig:PBC 16x12} in the Supplementary Materials), indicating the crossover is a finite size and open boundary effect. The ability to handle systems under PBC with NQS can mitigate the finite size effect, allowing to obtain more reliable results in the thermodynamic limit. By studying systems under PBC, we observe a robust, filled stripe phase with wavelength of $\lambda = 8$ as shown in Figure~\ref{fig:PBC 16x16, PBC t2 16x12, 32x8 Combined}(A), a result that is consistent with previous findings~\cite{science.aam7127}. Simulations on smaller $8\times 8$ and $12\times 12$ lattices (see Figures~\ref{fig:PBC 8x8}-\ref{fig:PBC 12x12} in the Supplementary Materials) show clear finite-size effects due to their incommensurability with the wavelength $\lambda=8$.

\subsection*{Half-filled stripe order in $t^\prime$-Hubbard model}

It was found that the doped pure Hubbard model doesn't host superconductivity \cite{qin2020absence}. To pursue superconducting ground states, the simplest modification is to include the next nearest neighboring hopping ($t^\prime$) term. On one hand, a finite $t^\prime$ can account for the electron and hole doping asymmetry in the phase diagram of cuprates. On the other hand, electronic structure calculations \cite{andersen1995lda,PhysRevB.98.134501} show a finite $t^\prime \approx-0.2$ term indeed exists in the one band effective Hamiltonian of cuprates. A recent study by Xu et al.~\cite{xu2024coexistence} found that the phase diagram of the Hubbard model with $t^\prime =-0.2$ agrees well with the cuprates, where fractional filling stripe coexists with the d-wave superconductivity in the hole doping region. The $t^\prime$ term frustrates the anti-ferromagnetic order in the parent compound and makes the doped stripe state fragile. Importantly, large finite size and boundary effect \cite{PhysRevB.102.041106} was found for the wavelength of the stripe state in the hole doping region, which fluctuates in the vicinity of $\lambda = 4$ \cite{xu2024coexistence}.

We perform systematic calculations for systems under PBC with $t^\prime = -0.2$. With PBC, we can get rid of the boundary effect. Our calculations on large sizes show the ground state for the $t^\prime = -0.2$ Hubbard model is clearly the half-filled stripe, i.e., with wavelength $\lambda =4$, as shown in Figure \ref{fig:PBC 16x16, PBC t2 16x12, 32x8 Combined}(B). These results indicate that the previous found fluctuating wavelengths for stripe may result from the finite size and boundary conditions. The $\lambda = 4$ stripe agrees well with the experimental observation of cuprates \cite{tranquada1995evidence}, providing further evidence to support the relevance of the $t^\prime$ Hubbard model to cuprates.

\subsubsection*{The direction of the stripe state}

\begin{figure} %
	\centering
	\includegraphics[width=0.6\textwidth]{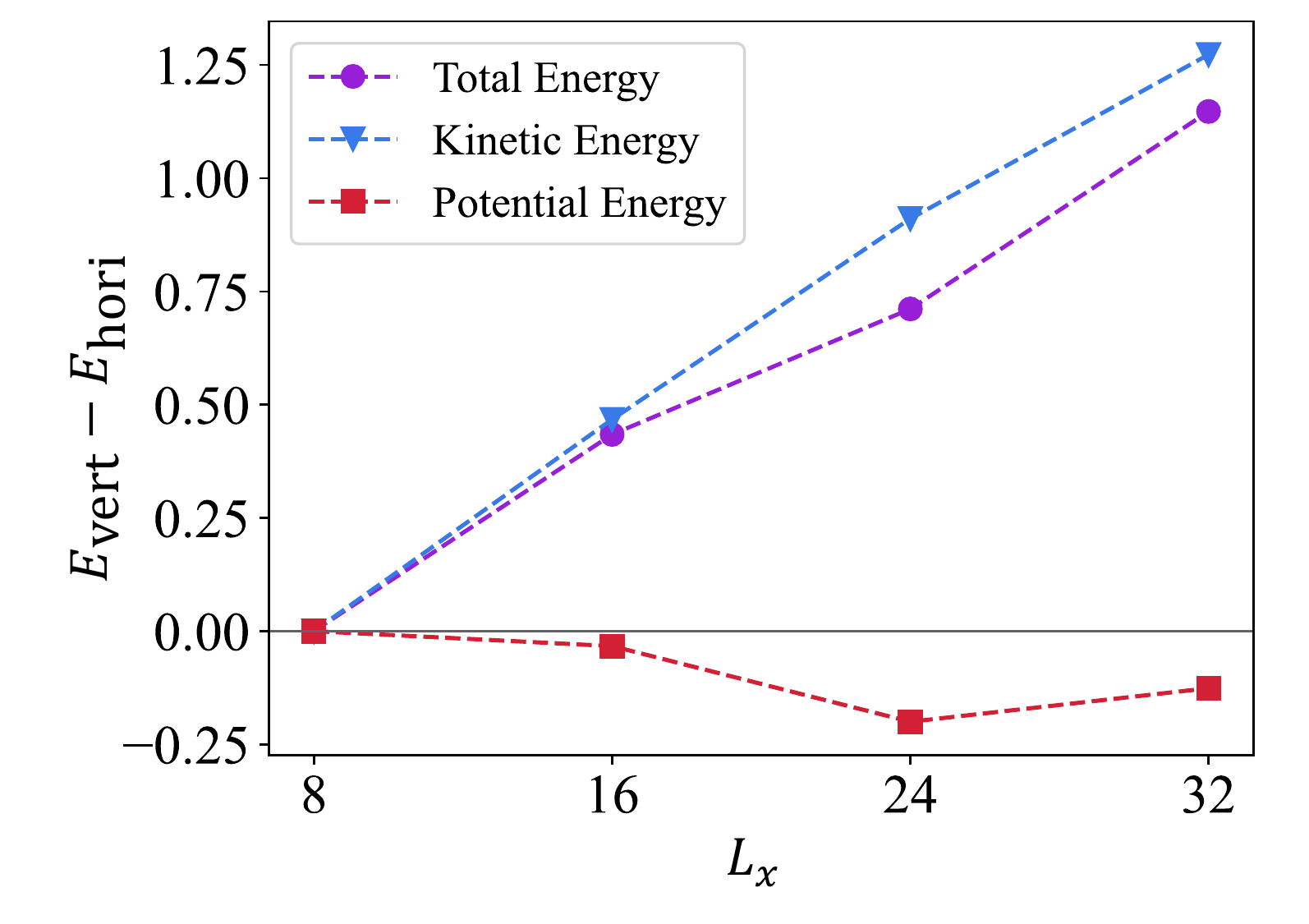} %

	\caption{\textbf{Energy difference between vertical stripe and horizontal stripe.} Kinetic energy, potential energy, and total energy results are all shown. We systematically study systems $L_x \times 8$ with $L_x$ ranging from 8 to 32, with $t^\prime=-0.2$ and $\delta =1/8$. We can find that the horizontal stripe has lower energy for all the systems (notice that for $8 \times 8$ system, vertical and horizontal stripes are the same). Moreover, the results show that the gain of energy in the horizontal stripe is mainly from the kinetic energy.}
	\label{fig:vert_hori} %
\end{figure}

Previous numerical studies on stripe order in the Hubbard and related models have been conducted on cylindrical geometries with DMRG \cite{PhysRevLett.80.1272,PhysRevResearch.2.033073,2018npjQM...3...22H,science.aal5304,PhysRevLett.132.066002}. However, such boundary conditions and their related finite-size effect can inherently favor the formation of vertical stripe orientations. 
To alleviate the possible boundary effect, we study systems with PBC in both directions. We systematically study systems with size $L_x \times 8$ where $L_x$ is ranged from $8$ to $32$, with $t^\prime=-0.2$ and $\delta =1/8$ (see Equation \ref{Hamil}). We consistently observe that a horizontal stripe configuration is energetically favored in the ground state while the system can accommodate half-filled stripe in both horizontal and vertical directions (see Figure~\ref{fig:PBC 16x16, PBC t2 16x12, 32x8 Combined}(B) and (C), Figure~\ref{fig:vert_hori} and~\ref{fig:PBC t2 8x8}-\ref{fig:PBC t2 24x8} in the Supplementary Materials).
These results suggest the existence of a tendency toward fewer long stripes rather than multiple fragmented ones in the systems, different from previous studies on cylinders \cite{xu2024coexistence}. From Figure~\ref{fig:vert_hori}, we can also find that the gain of energy for the horizontal stripe state is mainly from the kinetic energy, while the potential energy in the horizontal stripe is slightly higher. A simple explanation is that the formation of a long ``river" of holes in the horizontal stripe is beneficial to the kinetic energy. A comprehensive understanding of this observation needs further investigation.

For rectangular systems with a width not divisible by $8$, a horizontal stripe with wavelength $\lambda = 4$ is impossible, so ``frustration" appears. One such example is the $16\times10$ system, for which we find that the trend for horizontal stripe wins and the ground state has a wavelength of $\lambda = 5$ while the vertical stripe state with $\lambda = 4$ has a higher energy (see Figure~\ref{fig:PBC t2 16x10}-\ref{fig:PBC t2 16x10 new} in the Supplementary Materials). In another example for the  $16\times12$ system, the trend for wavelength $4$ wins in the competition and the ground state is a vertical stripe (see Figure~\ref{fig:PBC t2 16x12} in the Supplementary Materials). 

For the pure Hubbard model, it is also possible that the true ground state is a horizontal stripe. But the resolution of this issue requires the study of systems that can accommodate stripes with a wavelength $8$ in both vertical and horizontal directions. The smallest size for such systems under PBC is $32 \times 16$, which is beyond our present capacity.

\subsection*{Conclusions and outlook}

In this work, we propose a new NQS framework for studying quantum many-body systems by leveraging the cutting-edge machine learning architectures and
developing highly accurate optimization algorithms. State-of-the-art results for the 2D doped Hubbard model with unprecedented sizes are achieved with this new framework. NQS enables the study of 2D Hubbard model under periodic boundary conditions, which is crucial to get rid of the boundary effect and makes the extrapolation to the thermodynamic limit reliable. By studying 2D Hubbard model with large systems under periodic boundary conditions, we confirm the filled stripe phase in the ground state of the doped pure ($t^\prime = 0$) Hubbard model. When the next nearest neighboring hopping term ($t^\prime = -0.2$) is included, the ground state is found to be half-filled stripes, consistent with experimental observation in cuprates. Noticeably, the direction of the stripe for $t^\prime = -0.2$ is determined to be horizontal when the width and the length are not equal, different from previous studies \cite{xu2024coexistence}. These results indicate that previous calculations on systems with open boundary suffer from the boundary effect and underscore the effectiveness of the new NQS method in gaining more reliable results. 

We also have a careful analysis of the attention patterns learned in the ground state of 2D Hubbard model. Interestingly, we find that after the energy optimizations, different heads within our NQS specialize in encoding the correlations at different scales of the ground-state wave function. This special structure enables our NQS ansatz to effectively capture long-range correlations and entanglements, which is crucial for the accurate calculation of strongly correlated quantum many-body systems.

There is much room for the future development of NQS. On one hand, the development of more efficient optimization approaches could enable the study of even large system sizes, which is crucial for more reliable finite size scaling in the doped Hubbard model. On the other hand, more sophisticated NQS ansatz for the calculation of dynamic and finite temperature properties \cite{doi:10.1126/science.ade9194} are needed to further unveil the physics of the Hubbard model and establish the microscopic theory for cuprate. We anticipate that NQS framework will continue benefiting from the development in the broad machine learning community. Nevertheless, our results demonstrate that NQS is already a powerful framework for solving challenging many-fermions systems.

\subsection*{Methods summary}

In VMC, we employ Markov chain Monte Carlo (MCMC) to sample electron configurations according to $\psi_{\theta}^2(\mn)$. Following convergence, a separate inference run is performed with the fixed, optimized NQS to compute the physical observables until the desired statistical accuracy is achieved. Further details on the network architecture, hyperparameters, and algorithmic specifics are provided in the supplementary materials.

\clearpage %

\bibliography{science_template} %
\bibliographystyle{sciencemag}

\section*{Acknowledgements}
We thank ByteDance Seed for inspiration and encouragement, and Hang Li for his guidance and support. We thank Xiangjian Qian for providing DMRG benchmark results. We are grateful to Yuxi Zhu for helping with graphics.

\paragraph*{Funding:}
D.H. is supported by National Science Foundation of China (NSFC62376007).
Y.W. is supported by a start-up grant from IOP-CAS.
T.X. is supported by National Natural Science Foundation of China (Grant No. 12488201).
M.Q. acknowledges the support from the National Key Research and Development Program of MOST of China (2022YFA1405400), the National Natural Science Foundation of China (Grant No. 12274290) and the Innovation Program for Quantum Science and Technology (2021ZD0301902).
L.W. is supported by National Science and Technology Major Project (2022ZD0114902) and National Science Foundation of China (NSFC92470123, NSFC62276005).

\paragraph*{Author contributions:}
L.W. and D.L. conceived this project. Y.G. designed the algorithm, wrote the code and performed the calculation with helps from  W.L., H.L., R.L and Y.H, all of whom contributed to preliminary explorations. Y.G. and W.L. visualized the data. B.Z., Y.W., M.Q. and T.X. provided valuable insight in the physics of the Hubbard model. Y.W. wrote the PEPS code for the benchmark with $t'$ hopping. D.H. provided insights into neural network model design.  All authors contributed to the analysis of the data, the discussion of the results and the writing of the manuscript. 

\paragraph*{Competing interests:}

There are no competing interests to declare.

\paragraph*{Data and materials availability:}

All code and data will be made openly available upon publication.

\subsection*{Supplementary materials}
Supplementary Text\\
Figures S1 to S34\\
Tables S1 to S7

\newpage

\renewcommand{\thefigure}{S\arabic{figure}}
\renewcommand{\thetable}{S\arabic{table}}
\renewcommand{\theequation}{S\arabic{equation}}
\renewcommand{\thepage}{S\arabic{page}}
\setcounter{figure}{0}
\setcounter{table}{0}
\setcounter{equation}{0}
\setcounter{page}{1} %

\begin{center}
\section*{Supplementary Materials for\\
Solving the Hubbard model with Neural Quantum States}

\author{
	Yuntian Gu$^{\dagger}$,
	Wenrui Li$^{\dagger}$,
        Heng Lin$^{\dagger}$,
	Bo Zhan$^{\dagger}$,
        Ruichen Li,
        Yifei Huang,\\
        Di He$^{\ast}$, %
        Yantao Wu$^{\ast}$, %
        Tao Xiang$^{\ast}$, %
        Mingpu Qin$^{\ast}$, %
        Liwei Wang$^{\ast}$, %
        Dingshun Lv$^{\ast}$\\ %
	\small$^\ast$Corresponding author: dihe@pku.edu.cn (D. He);  yantaow@iphy.ac.cn (Y. T. Wu);\\
    \small txiang@iphy.ac.cn (T. Xiang); qinmingpu@sjtu.edu.cn(M. P. Qin); \\
    \small wanglw@pku.edu.cn (L. W. Wang); lvdingshun@bytedance.com (D. S. Lv).\\
     \small$^\dagger$These authors contributed equally to this work.
}
\end{center}

\subsubsection*{This PDF file includes:}
Supplementary Text\\
Figures S1 to S34\\
Tables S1 to S7

\newpage

\section{Variational Monte Carlo}

VMC is a quantum Monte Carlo method based on the variational principle of quantum mechanics. Upon choosing a wavefunction ansatz $\psi_{\theta}$, minimizing the energy of this ansatz with respect to the parameters $\theta$ makes $\psi_{\theta}$ an approximation to the ground state. 
In this work, we adopt a neural network as the wavefunction ansatz. The neural network takes occupation number $\mn$ as input and outputs the amplitude of the wavefunction, $\braket{\mn|\psi_{\theta}}=\psi_{\theta}(\mn)$. 
The energy of the wavefunction can be written as the expectation value of the local energy, i.e.
\begin{equation}
\label{eqn:exact_energy}
    \bar{E}_{\theta} = \frac{\braket{\psi_{\theta} | \hat{H} | \psi_{\theta}}}{\braket{\psi_{\theta} | \psi_{\theta}}} = \mbE_{\mn \sim \psi_{\theta}^2 / \left\| \psi_{\theta} \right\|^2}[E_{\text{loc}}^{\theta}(\mn)],
\end{equation}
where 
\begin{equation}
    \label{eqn:E_local}
    E_{\text{loc}}^{\theta}(\mn) = \sum\limits_{\mn ^\prime}\frac{\psi_{\theta}(\mn ^\prime)}{\psi_{\theta}(\mn)}\Bra{\mn ^\prime}\hat{H}\Ket{\mn}
\end{equation} 
is the local energy.
Exact calculation of the energy using Equation~\ref{eqn:exact_energy} is infeasible in an exponentially large Hilbert space. However, we can approximate this summation using the Monte Carlo method.
A set of occupation numbers $\{\mn _i \}_{i=1}^{B}$ is sampled from the probability distribution defined by the squared amplitude of the wavefunction using Markov chain Monte Carlo (MCMC) techniques.
The local energy of occupation numbers, $E_{\text{loc}}^\theta(\mn)$, can be efficiently calculated from the Hubbard model Hamiltonian, with the number of summation terms in Equation~\ref{eqn:E_local} scaling linearly with the size of the system. Therefore, parameters of the neural network can be optimized to approachthe  ground state of the Hubbard model: 
\begin{equation}
    \theta^* = \arg \min_{\theta} \bar{E}_{\theta}.
\end{equation}

\subsection{Neural Quantum State Wavefunction Ansatz}
As illustrated in Figure~\ref{fig:arc}, we employ a neural quantum state (NQS) ansatz to represent the many-body quantum state $\ket{\psi}$ of a fermionic system with $N$ lattice sites. Each individual site corresponds to a $4$ dimensional physical Hilbert space spanned by $\ket{0}$, 
 $\ket{\uparrow}$, $\ket{\downarrow}$, and $\ket{\uparrow \downarrow}$. We encode the four possible tokens into a $d$ dimensional vector via an embedding matrix $E \in \mathbb{R} ^{4 \times d}$. To account for spatial arrangement, crucial due to the permutation equivariance of self-attention, these embeddings are augmented with learnable positional encodings. To this end, we add the matrix $P \in \mathbb{R} ^{N \times d}$ to the input embedding. The embedded input can be compactly written into a matrix $X^{(0)} = [E(n_1) + P_1, ..., E(n_N) + P_N]^\top \in \mathbb{R}^{N \times d}$. Then, $L$ Transformer blocks follow, each of which transforms the input via the following equation:
\begin{align}
Y^{(l)} &= X^{(l-1)} + \text{Attn}^{(l)} (X^{(l-1)}), \\
X^{(l)} &= Y^{(l)} + \text{FFN}^{(l)}  (Y^{(l)}),
\end{align}
where $l \in [L]$, $\text{Attn}^{(l)}$ and $\text{FFN}^{(l)}$ denote the multi-head self-attention layer and the feed-forward network for the $l$-th Transformer block, respectively:
\begin{align}
\label{eq:transformer}
    \text{Attn}^{(l)}(X)&=\sum_{h=1}^H \text{softmax}\left(X W_Q^{(l,h)}(X W_K^{(l,h)})^\top / \sqrt{d_H}\right)XW_V^{(l,h)}(W_O^{(l,h)})^\top,\\
    \text{FFN}^{(l)}(X)&=\sigma(XW_F^{(l)}).
\end{align}
The variational parameters $W_Q^{(l,h)}, W_K^{(l,h)}, W_V^{(l,h)}, W_O^{(l,h)} \in \mathbb{R}^{d\times d _ H}$ are the query, key, value, and output matrices of the head $h$, respectively. $d_H = d / H$ is the hidden dimension of each head. $W_F^{(l)} \in \mathbb{R}^{d\times d}$ is the weight matrix in the FFN$^{(l)}$. The activation $\sigma$ is chosen as SiLU~\cite{elfwing2018sigmoid}.

The final output $X^{(L)}$ is transformered by a linear layer to produce $K$ sets of backflow orbitals $M ^{k} \in \mathbb{R}^{2N \times N_e}$, where $k \in [K]$. The Slater matrix $\Phi ^k$ is formed by selecting rows of $M^{k}$ corresponding to the occupied position in the input configuration $\mn$. The overall NQS wavefunction is a linear combination of these determinants:
\begin{equation}
\braket{\mn |\psi} = \sum _{k=1}^K \text{det}[\Phi ^k].
\end{equation}

\subsection{Optimization}
Stochastic reconfiguration (SR) is one of the most popular optimization methods for NQS. In this work, however, we propose an advanced SR algorithms for superior performance, named Moment-Adaptive ReConfiguration Heuristic (MARCH), most directly inspired by the SPRING algorithm~\cite{goldshlager2024kaczmarz}. MARCH combines ideas from the extremely successful Adam optimizer in traditional machine learning~\cite{kingma2017adammethodstochasticoptimization} and the recently introduced efficient Min-SR algorithm~\cite{chen2024empowering, rende2024simple}.%

In SR, we optimize variational wavefunction $\ket{\psi_{\theta}}$ by finding parameter updates that best approach imaginary-time evolution wavefunction $e^{-\tau\hat{H}}\ket{\psi}$. As imaginary time evolution may push the state out of the parameter manifold, 
SR utilizes the Fubini-Study (FS) distance to measure the quantum distance and pull the imaginary-time evolving state back. In general, given two quantum states $\ket{\psi},\ket{\phi}$, FS distance is defined as follows:
\begin{equation}
    d(\ket{\psi},\ket{\phi})=\text{arccos}\frac{\left|\braket{\psi|\phi}\right|}{||\psi|| \cdot ||\phi||}.
\end{equation}

By expanding the FS distance to the lowest order, one derives:
\begin{equation}
    d(\ket{\psi}+\ket{\delta\psi},\ket{\psi}+\ket{\delta\phi})=\left|\left|\left(\frac{\ket{\delta\psi}}{||\psi||}- \frac{\ket{\delta\phi}}{||\psi||}\right)-\left(\frac{\braket{\psi|\delta\psi}\ket{\psi}}{||\psi||^3}-\frac{\braket{\psi|\delta\phi}\ket{\psi}}{||\psi||^3}\right)\right|\right|.
\label{opt: main eq}
\end{equation}

To measure the FS distance between the variational state $\ket{\psi_{\theta+\delta\theta}}$ and the imaginary-time evolving state $\ket{\psi'}=e^{-\delta \tau \hat{H}}\ket{\psi_{\theta}}$, we first expand the quantum states to first order as follows:
\begin{equation}
    \ket{\delta\psi_{\theta}}=\sum\limits_{\mn}\sum\limits_{k}\frac{\partial\psi_{\theta}(\mn)}{\partial\theta_{k}}\delta\theta_k\ket{\mn}=\sum\limits_{\mn}\psi_{\theta}(\mn)\sum\limits_{k}\bar{O}_{k}(\mn)\delta\theta_k\ket{\mn},
\label{opt: eq1}
\end{equation}
\begin{equation}
    \ket{\delta\psi'}=-\hat{H}\delta\tau\ket{\psi_{\theta}}=-\delta\tau\sum\limits_{\mn}\psi_{\theta}(\mn)E_{\text{loc}}^\theta(\mn)\ket{\mn},
\label{opt: eq2}
\end{equation}
where $\bar{O}_{k}(\mn)=\frac{\partial}{\partial\theta_k}\text{log}\left|\psi_{\theta}(\mn)\right|$.

Putting Equation~\ref{opt: eq1} and Equation~\ref{opt: eq2} into Equation~\ref{opt: main eq}, we obtain:
\begin{equation}
d^2(\ket{\psi_\theta}+\ket{\delta\psi_{\theta}},\ket{\psi_\theta}+\ket{\delta\psi'})
    =\sum\limits_{\mn}\frac{\psi^2_\theta(\mn)}{||\psi_{\theta}||^2}\left|\sum\limits_{k}O_{k}(\mn)\cdot \delta\theta_k-\epsilon(\mn)\right|^2,
\end{equation}
where $O_k(\mn)=\bar{O}_k(\mn)-\mbE_{\mn '} [\bar{O}_k(\mn ')], O(\mn) = \left(O_1(\mn),\cdots,O_k(\mn),\cdots\right)^T, \epsilon (\mn) = -\delta\tau (E_{\text{loc}}(\mn) - \mbE_{\mn '} [E_{\text{loc}}(\mn ')])$.

Therefore, the SR parameter update is typically defined as 
\begin{equation}
    \dd \theta = \argmin _{\dd \theta '} \mbE_\mn \left[|O(\mn)\dd \theta ' - \epsilon (\mn)|^2\right]\approx \argmin _{\dd \theta '} \frac{1}{B}\sum\limits_{i=1}^{B} |O(\mn_i)\dd \theta ' - \epsilon (\mn_i)|^2,
\end{equation}
where $B$ denotes batch size in MCMC sampling.
 
To express the problem in a standard linear algebra form, we construct a matrix $\tilde{O}$ and a vector $\tilde{\epsilon}$ by stacking the contributions from each of the $B$ samples in the batch:
\begin{equation}
    \tilde{O} = \begin{pmatrix}
 O(\mn_1) \\
 O(\mn_2) \\
 \vdots \\
 O(\mn_B)
\end{pmatrix}
\in \mathbb{R}^{B \times N_{\text{params}}}
\quad \text{and} \quad
\tilde{\epsilon} = \begin{pmatrix}
 \epsilon(\mn_1) \\
 \epsilon(\mn_2) \\
 \vdots \\
 \epsilon(\mn_B)
\end{pmatrix}
\in \mathbb{R}^{B}.
\end{equation}

Since $\tilde O$ might be ill-conditioned, the conventional method is to add some form of regularization. For example, a Tikhonov regularization is usually added to yield the regularized problem as follows:
\begin{equation}
    \dd \theta = \argmin _{\dd \theta '} \frac{1}{\lambda}||\tilde O\dd \theta ' - \tilde{\epsilon}||^2 + ||\dd \theta '||^2.
\end{equation}

MARCH differs from previous SR algorithms by adaptively translating and scaling $\dd \theta$ individually for each parameter based on the estimation of the first and second moments.

The overall objective function can be written as:
\begin{equation}
\dd \theta_k = \argmin _{\dd \theta '} \frac{1}{\lambda}||\tilde{O}\dd \theta ' - \tilde{\epsilon}||^2 + ||\text{diag}(v_{k-1})^{1/4}(\dd \theta' - \phi_{k-1})||^2,
\label{eq:optimization}
\end{equation}
where $k$ denotes the current step, $\text{diag}(v_{k-1})$ is the diagonal matrix whose main diagonal is composed of $v_{k-1}$. The first-order moment estimator, $\phi_{k} = \mu \dd \theta _{k}$, acts as a momentum term. Intuitively, $\phi$ carries the momentum, which accelerates updates in directions of persistent gradients and dampens oscillations. The second moment estimator, 
$v_k = \beta v_{k-1} + (\dd \theta _k - \dd \theta _{k-1})^2$
, tracks the volatility of the gradient. A large $v_k$ indicates that the gradient is changing erratically, so the update of the parameter is suppressed to prevent divergence. Conversely, a small $v_k$ implies a stable gradient, allowing the parameter to be updated more aggressively. The behavior of these estimators is governed by the hyperparameters $\mu$, $\beta$, and $\lambda$, which are specified in Table \ref{tab:transformer}.

Therefore, the explicit update formula can be derived. We first define $\pi = \text{diag}(v_{k-1})^{1/4} (\dd \theta ' - \phi_{k-1}), U=\tilde{O}\text{diag} (v_{k-1})^{-1/4}, \zeta = \tilde{\epsilon} - \tilde{O}\phi _{k-1}$. The equation~\ref{eq:optimization} can be recast as:
\begin{equation}
\pi = \argmin _{\pi '} \frac{1}{\lambda}||U\pi '  - \zeta||^2 + ||\pi '||^2.
\end{equation}

We can reduce the computational budget from $\mathcal{O}(N_{\text{params}}^3)$ to $\mathcal{O}(B^2 N_{\text{params}})$ by Woodbury matrix identity:
\begin{align}
    \pi &= (U^TU + \lambda I)^{-1} U^T \zeta\\
    &= \frac{1}{\lambda} \left[I - U^T(\lambda I + UU^T)^{-1} U \right] U^T\zeta \\
    &= \frac{1}{\lambda} \left[U^T - U^T(\lambda I + UU^T)^{-1} (U U^T + \lambda I - \lambda I)\right]\zeta \\
    &= U^T(\lambda I + UU^T)^{-1} \zeta.
\end{align}

Finally, we can write down the parameters update formula as:
\begin{equation}
\dd\theta = \text{diag}(v_{k-1})^{-1/2}\tilde{O}^T(\tilde{O} \text{diag}(v_{k-1})^{-1/2} \tilde{O}^T + \lambda I)^{-1}(\tilde{\epsilon} - \tilde{O}\phi _{k-1}) + \phi _{k-1}.
\end{equation} 

\section{Pretrain}

Instead of randomly initializing the NQS ansatz, we use a pretrain method to provide the transformer wavefunction better initial parameters before optimizing it with MARCH.

We first train a simple neural network similar to neural network backflow (NNB)~\cite{luo2019backflow}. A multilayer perceptron (MLP) with two hidden layers maps the occupation number $\mn \in \{0, 1\}^{2N}$ to a orbital matrix $M_{\text{nnb}}(\mn) \in \mathbb{R}^{2N\times N_e}$, which is restricted to a block-diagonal matrix, 
\begin{equation}
M_{\text{nnb}}(\mn) = 
\begin{pmatrix}
    M_{\text{nnb}}^{\uparrow}(\mn) & 0 \\
    0 & M_{\text{nnb}}^{\downarrow}(\mn)
\end{pmatrix},
\end{equation}
where the $M_{\text{nnb}}^{\uparrow}(\mn), M_{\text{nnb}}^{\downarrow}(\mn)\in \mathbb{R}^{N\times N_e / 2}$ are the orbitals of spin-up and spin-down electrons, respectively. Based on the occupation number, $N_e$ rows of orbital matrix $M_{\text{nnb}}(\mn)$ are extracted to form a square orbital matrix $\Phi \in\mathbb{R}^{N_e\times N_e}$, which is also block-diagonal,
\begin{equation}
\Phi  = 
\begin{pmatrix}
    \Phi^{\uparrow} & 0 \\
    0 & \Phi^{\downarrow}
\end{pmatrix}.
\end{equation}
The amplitude of NNB ansatz is defined by the determinant of a matrix $\Phi$:
\begin{equation}
\psi_{\text{nnb}}(\mn) = \mathrm{det}[\Phi].
\end{equation}
This ansatz is variationally optimized using the VMC method, with the hyperparameters specified in Table~\ref{tab:nnb}. Subsequently, the optimized NNB ansatz is used to generate a labeled dataset, $S_{\text{pre}}$, of size $N_{\text{pre}}$:
\begin{equation}
S_{\text{pre}} = \left\{\mn_i,\ M _{\text{nnb}}(\mn_i)\right\}_{i=0}^{N_{\text{pre}}},
\end{equation}
where each entry consists of a configuration $\mn_i$ and its corresponding orbital matrix $M_{\text{nnb}}(\mn_i)$.

The transformer ansatz is then pretrained in a supervised manner, with the loss function to be the mean square error of the orbital matrix, i.e.,
\begin{equation}
    \mathcal{L}_{\text{pretrain}} = \frac{1}{N_{\text{pre}}}\sum_{i = 1}^{N_{\text{pre}}}\sum _{k=1}^K || M^{k}(\mn _i) - M_{\text{nnb}}(\mn_i) ||^2,
\end{equation}
with Adam optimizer. We note that in practice, the pretrain dataset can be generated online. Compared to random initialization, a pretrained transformer is much more stable and converges significantly faster.

\section{Pinning Field}

To avoid convergence to local minima during optimization, a temporary pinning field is applied. This field takes the form of an antiferromagnetic (AFM) term added to the Hamiltonian $\hat H$ on specific columns:

\begin{equation}
\hat H_{\text{pin}} = \hat H + h_m \sum_{i\in [L_x]} h_i\sum _{j\in [L_y]}(-1)^{j} \hat S _{ij}^z ,
\end{equation}
with $h_m=0.2$ and $h_i \in \{-1, 0, 1\}$ to probe the possible magnetic order. Initially, a transformer wavefunction $\psi_1$ is optimized under the influence of this pinning field. 
This pinned field guides the ansatz to fall into a desired pattern.
Subsequently, a second transformer ansatz $\psi_2$ is initialized. We then employ a pretraining procedure aimed at maximizing the fidelity $\frac{\braket{\psi_1 | \psi_2}^2}{\braket{\psi_1 | \psi_1}\braket{\psi_2 | \psi_2}}$. This effectively transfers the desired pattern to $\psi_2$. The final optimization is then performed on $\psi_2$ with the pinning field removed, allowing it to relax and find the true ground state.

With this method, we can stabilize a vertical or horizontal stripe. From Figure~\ref{fig:size_ablation} we can see that at the same network size, the vertical stripe is higher in energy than the horizontal stripe, indicating the horizontal stripe indeed is favored on $32\times 8$ lattice with $t'=-0.2$ and $\delta = 1/8$.

\section{Benchmark Results}

In addition to the results presented for the pure Hubbard model with OBC in the main text (numbers are listed in Table~\ref{tab:main}), this section details extensive benchmark comparisons of our NQS methodology against several leading numerical techniques across a variety of Hubbard model configurations and system parameters. We also provide an ablation study for our proposed optimizer MARCH against SPRING.

\subsection{Comparison with  Hidden Fermion Determinantal State}
We first benchmark our NQS approach against the Hidden Fermion Determinantal States (HFDS) method, a notable prior NQS implementation~\cite{robledo2022fermionic}. For the pure Hubbard model on a $16 \times 4$ lattice with PBC, which represents the largest calculation reported in their paper, our NQS yields a ground state energy of $-0.76298$. This value is significantly lower than the HFDS variational energy of $-0.753(2)$.

\subsection{Comparison with DMRG}
We further compare our NQS method with the DMRG on the $t'$ Hubbard model. Specifically, for a cylindrical geometry $16 \times 8$ with next-nearest-neighbor hopping $t^\prime = -0.2$, our NQS achieves a variational energy of $-0.72791$. This is lower than the DMRG energy of $-0.72747$, obtained with a large bond dimension of $m=40000$.

\subsection{Comparison with PEPS}

We also benchmark the computational efficiency of NQS against PEPS on identical hardware. The implementation of PEPS can be referred to Wu et al.~\cite{wu2025algorithmsvariationalmontecarlo}. As detailed in Figure~\ref{fig:gpu_peps}, the expressive power of NQS is similar to fermion PEPS with $D=10$. However, NQS is significantly better than PEPS in terms of computational efficiency under the same hardware conditions, as shown in Table~\ref{tab:speed}.

\subsection{Optimizer Ablation Study}
To assess the efficacy of our proposed optimizer MARCH, we performed an ablation study comparing its performance against SPRING~\cite{goldshlager2024kaczmarz} and Min-SR~\cite{chen2024empowering, rende2024simple}. The convergence behavior of both optimizers during the NQS training is presented in Figure~\ref{fig:optimizer_ablation}. These results clearly demonstrate that MARCH achieves significantly faster and more stable convergence compared to SPRING and Min-SR for the systems tested.

\subsection{Accuracy Estimation}

\label{sec:acc_estimate}

To compare the accuracy of different NQS, we benchmark them on relatively simple systems with known exact solutions. Specifically, we used a $4\times 4$ lattice or the half-filled pure Hubbard model, which can be solved exactly using Exact Diagonalization or AFQMC. The results are presented in Table~\ref{tab:estimate}.

\section{Hyperparameter}

We present general settings of NNB training, pretraining, and MARCH training in Table \ref{tab:nnb}, Table \ref{tab:pretrain}, and Table \ref{tab:transformer}, respectively. We note that for different systems, the hyperparameters might be slightly different. All code and data will be made openly available upon publication.

\begin{figure}
    \centering
    \includegraphics[width=\linewidth]{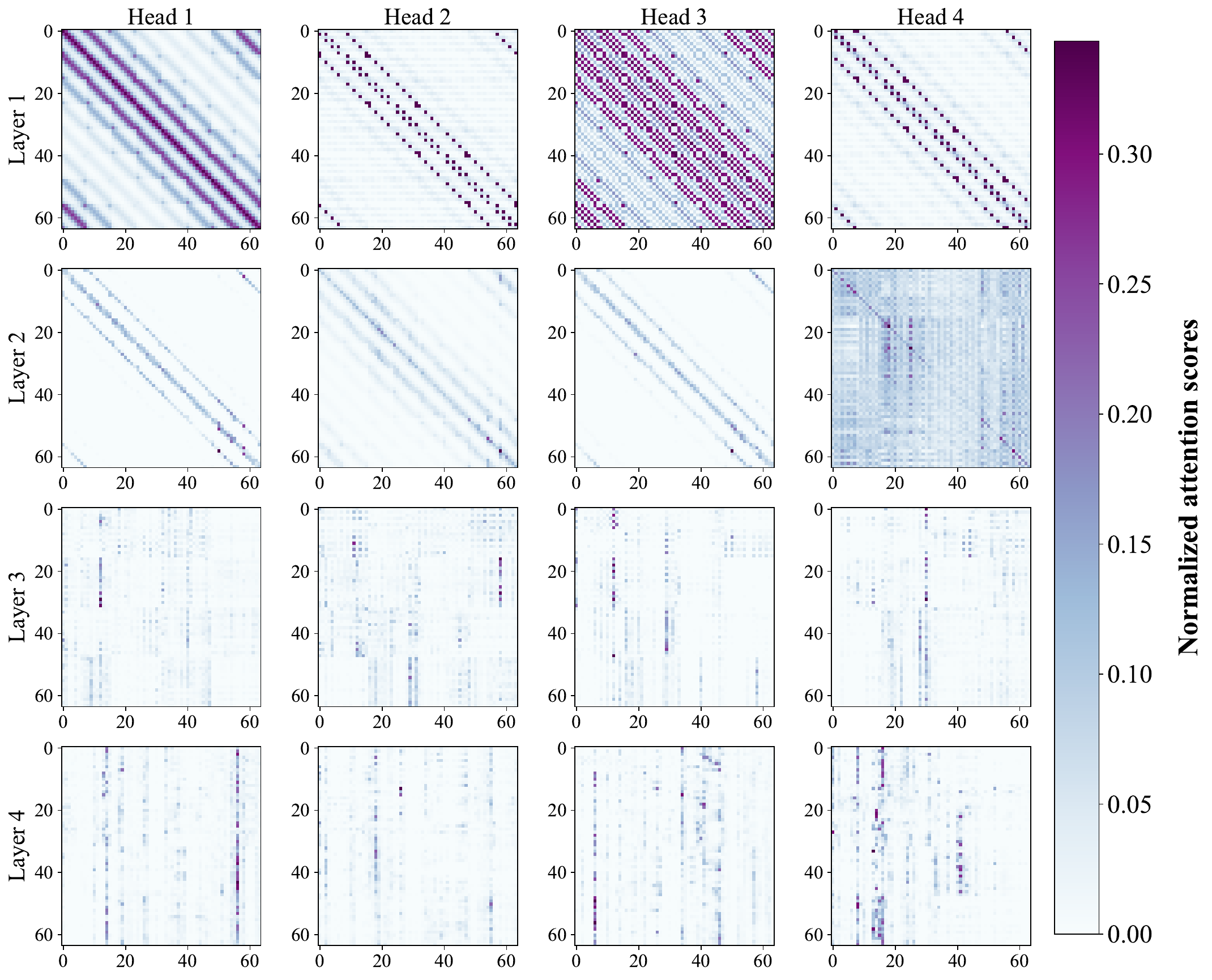}
    \caption{Attention maps for the transformer average over the sample electron configuration for $8\times 8$ pure Hubbard model at half-filling. The maps are organized by network layer (top to bottom: Layer 1 to Layer 4) and attention head (left to right: Head 1 to Head 4).}
    \label{fig:score 8x8 half}
\end{figure}

\begin{figure}
    \centering
    \includegraphics[width=\linewidth]{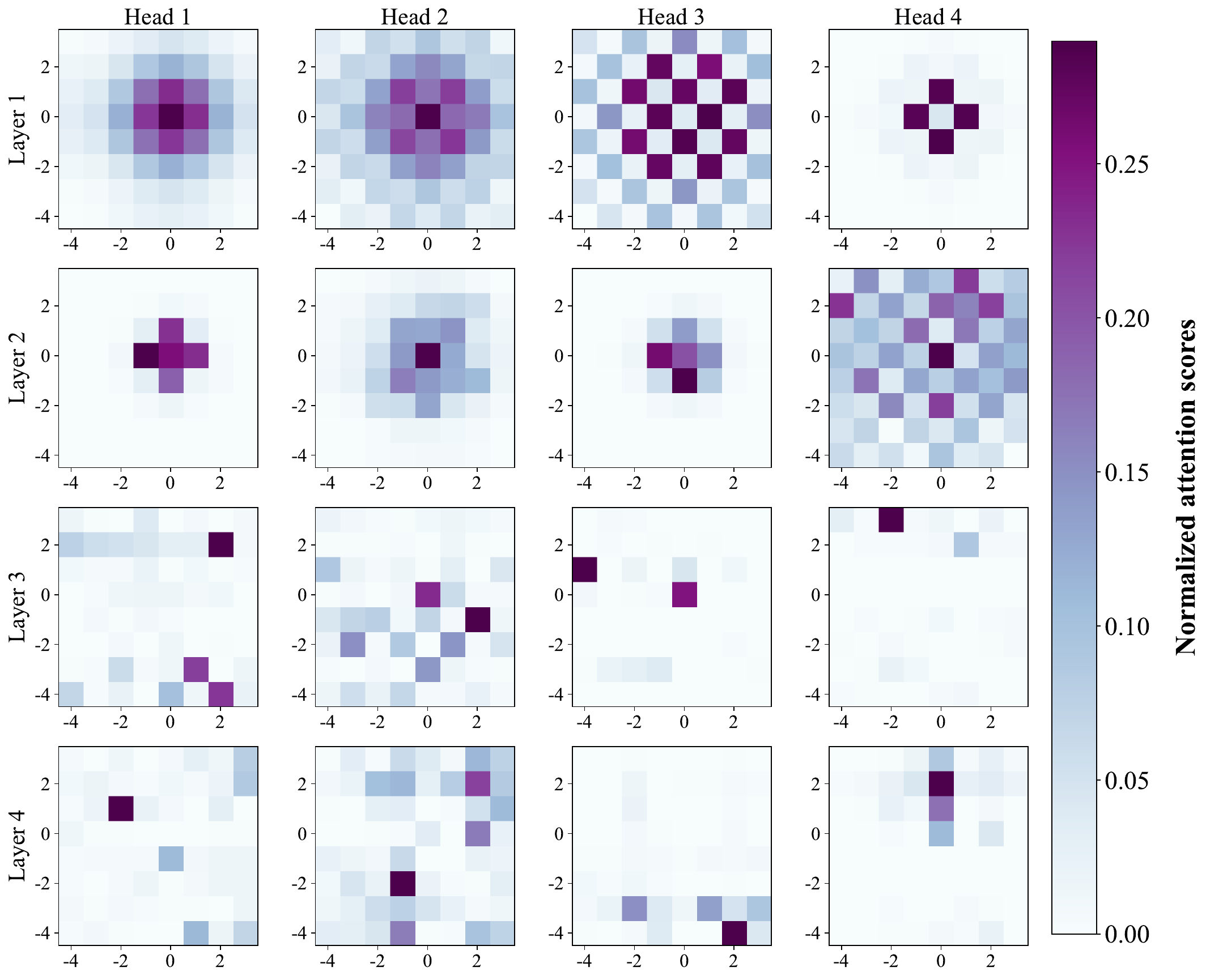}
    \caption{Visualization of the attention mechanism, showing how a lattice site attends to another target site for $8\times 8$ pure Hubbard model at half filling. The maps are organized by network layer (top to bottom: Layer 1 to Layer 4) and attention head (left to right: Head 1 to Head 4).}
    \label{fig:score 8x8 half site}
\end{figure}

\begin{figure}
    \centering
    \includegraphics[width=\linewidth]{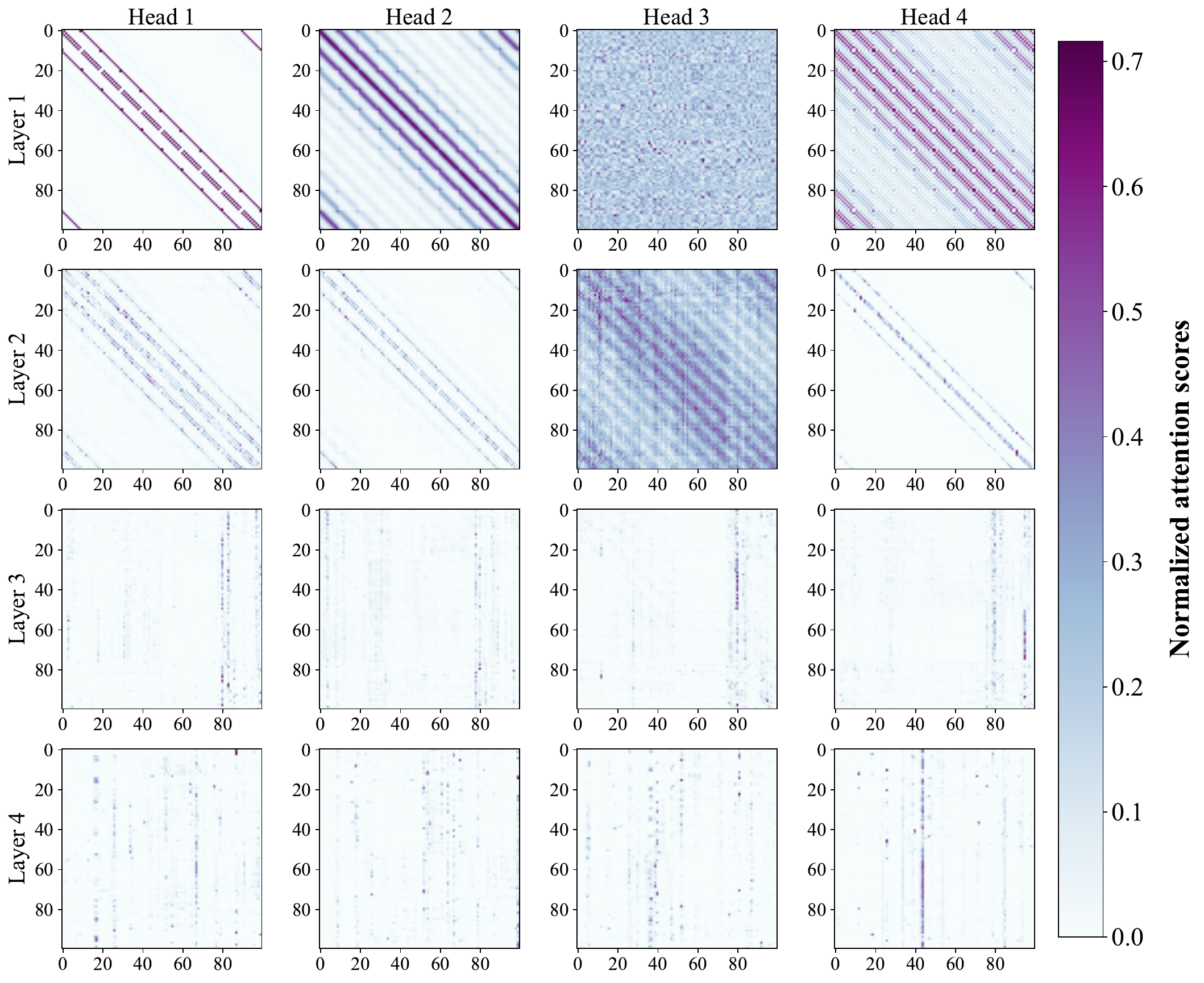}
    \caption{Similar as Figure~\ref{fig:score 8x8 half} but for $10 \times 10$ pure Hubbard model at half-filling.}
    \label{fig:score 10x10 half}
\end{figure}

\begin{figure}
    \centering
    \includegraphics[width=\linewidth]{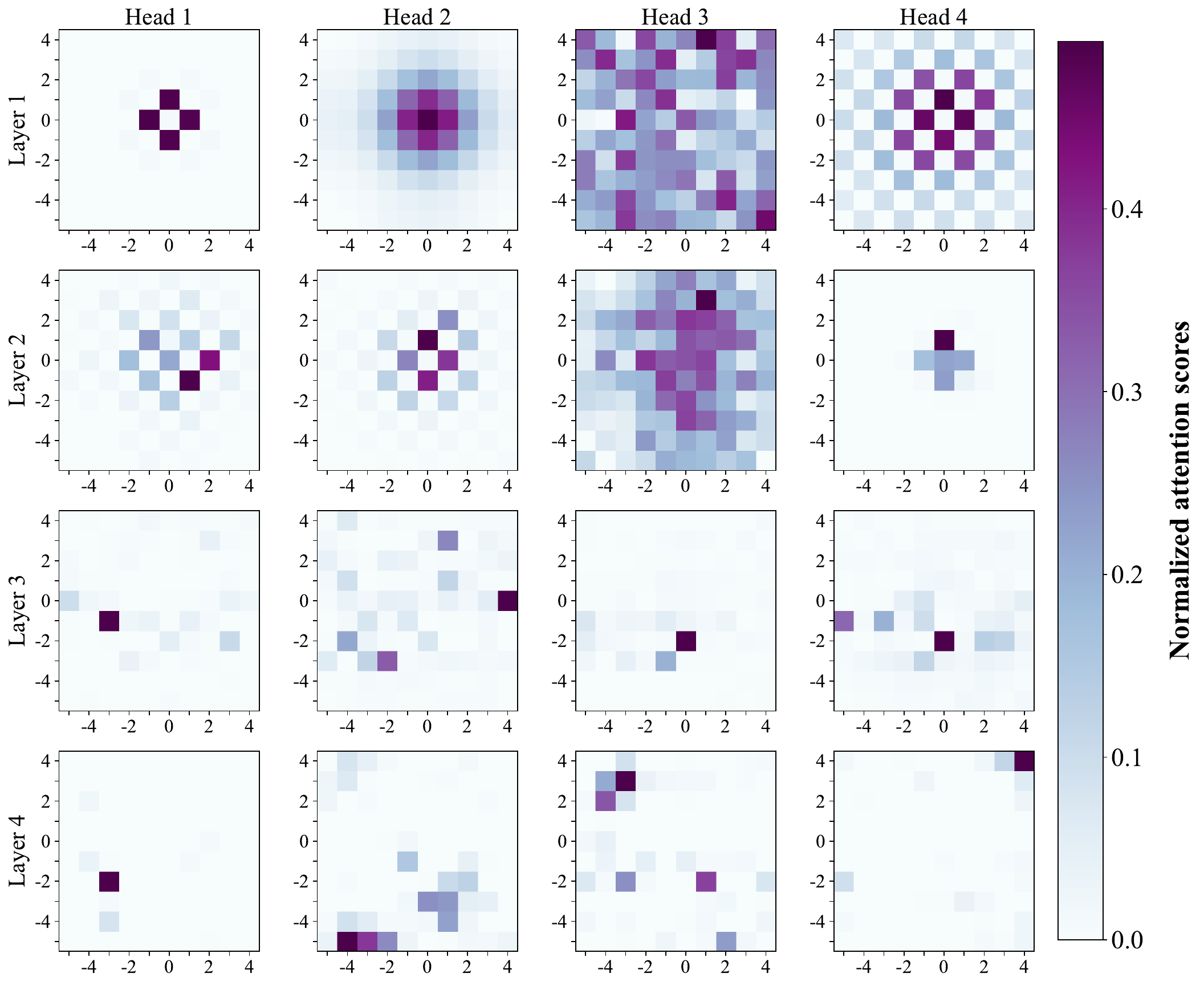}
    \caption{Similar as Figure~\ref{fig:score 8x8 half site} but for $10 \times 10$ pure Hubbard model at half-filling.}
    \label{fig:score 10x10 half site}
\end{figure}

\begin{figure}
    \centering
    \includegraphics[width=\linewidth]{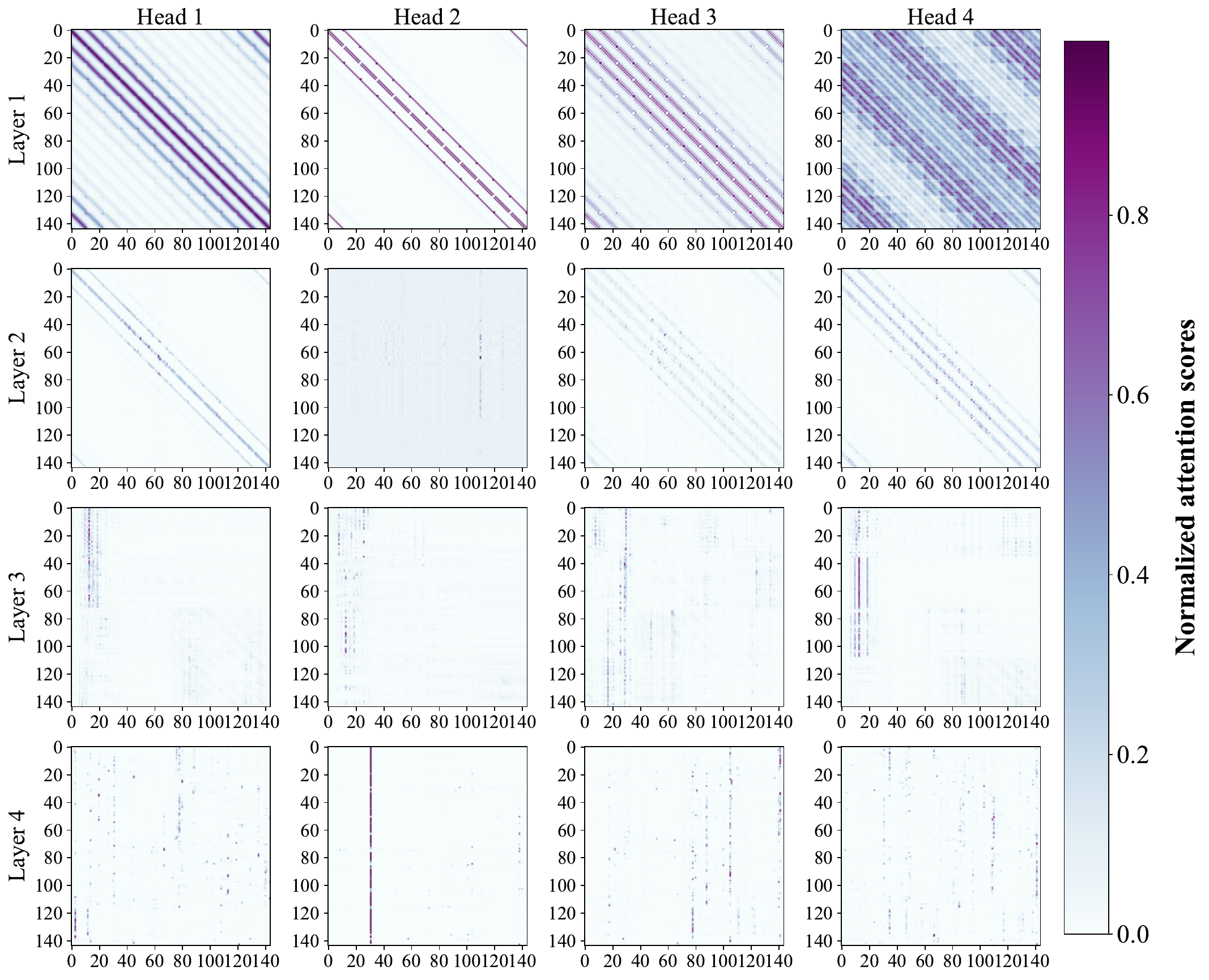}
    \caption{Similar as Figure~\ref{fig:score 8x8 half} but for $12 \times 12$ pure Hubbard model at half-filling.}
    \label{fig:score 12x12 half}
\end{figure}

\begin{figure}
    \centering
    \includegraphics[width=\linewidth]{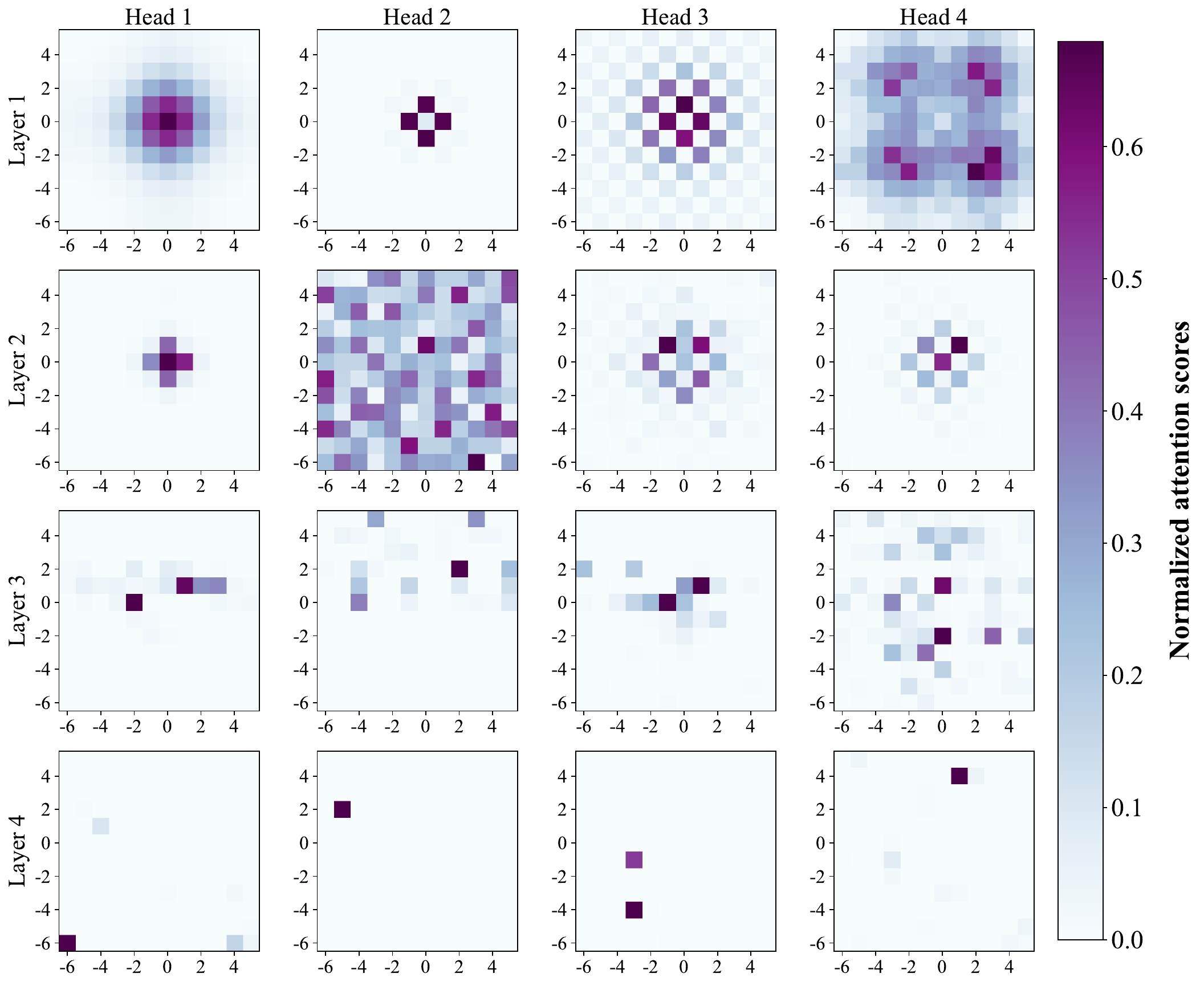}
    \caption{Similar as Figure~\ref{fig:score 8x8 half site} but for $12 \times 12$ pure Hubbard model at half-filling.}
    \label{fig:score 12x12 half site}
\end{figure}

\begin{figure}
    \centering
    \includegraphics[width=\linewidth]{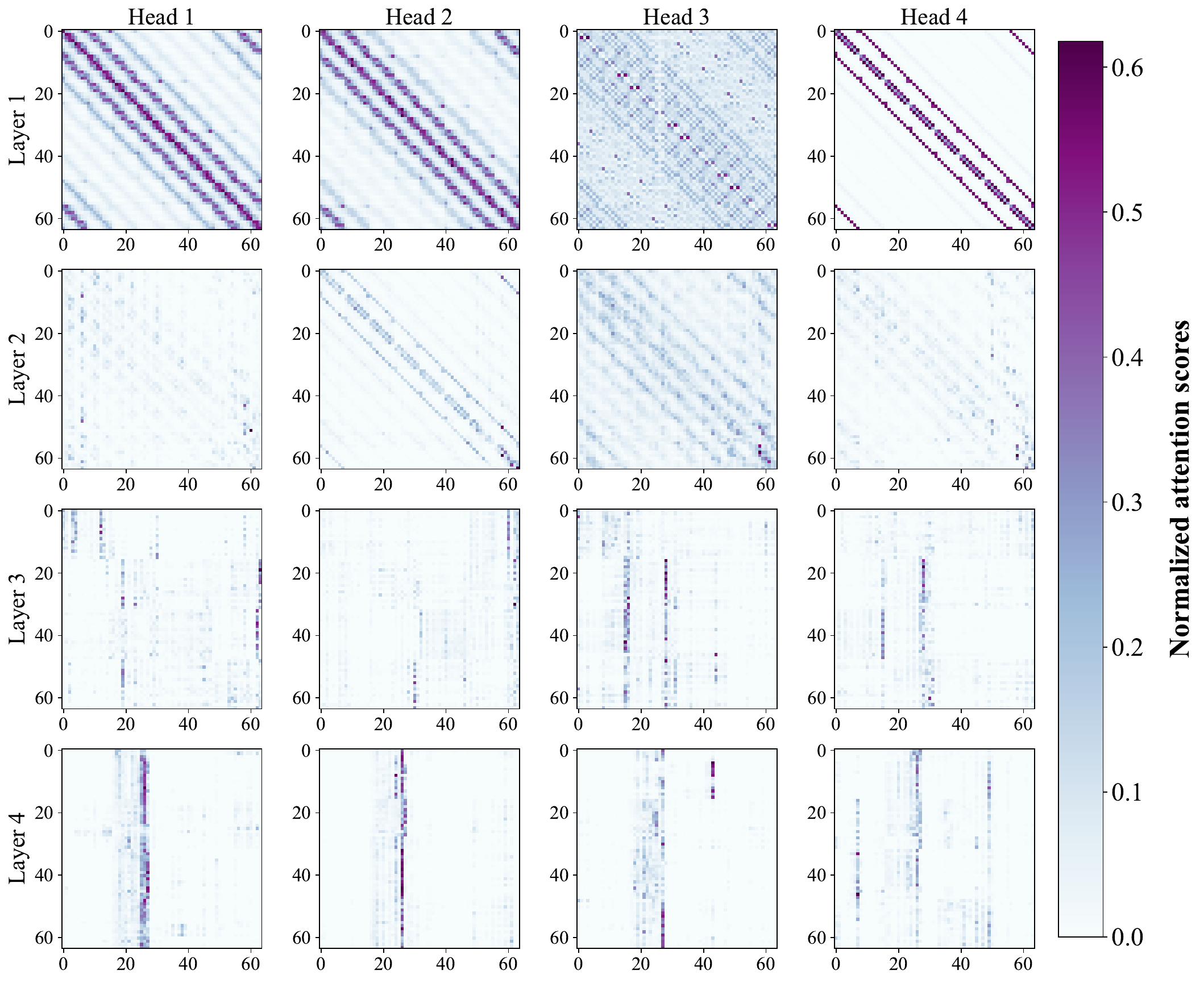}
    \caption{Similar as Figure~\ref{fig:score 8x8 half} but for $8 \times 8$ Hubbard model with $t' = -0.2$ at $1/8$ dopping.}
    \label{fig:score 8x8 doped}
\end{figure}

\begin{figure}
    \centering
    \includegraphics[width=\linewidth]{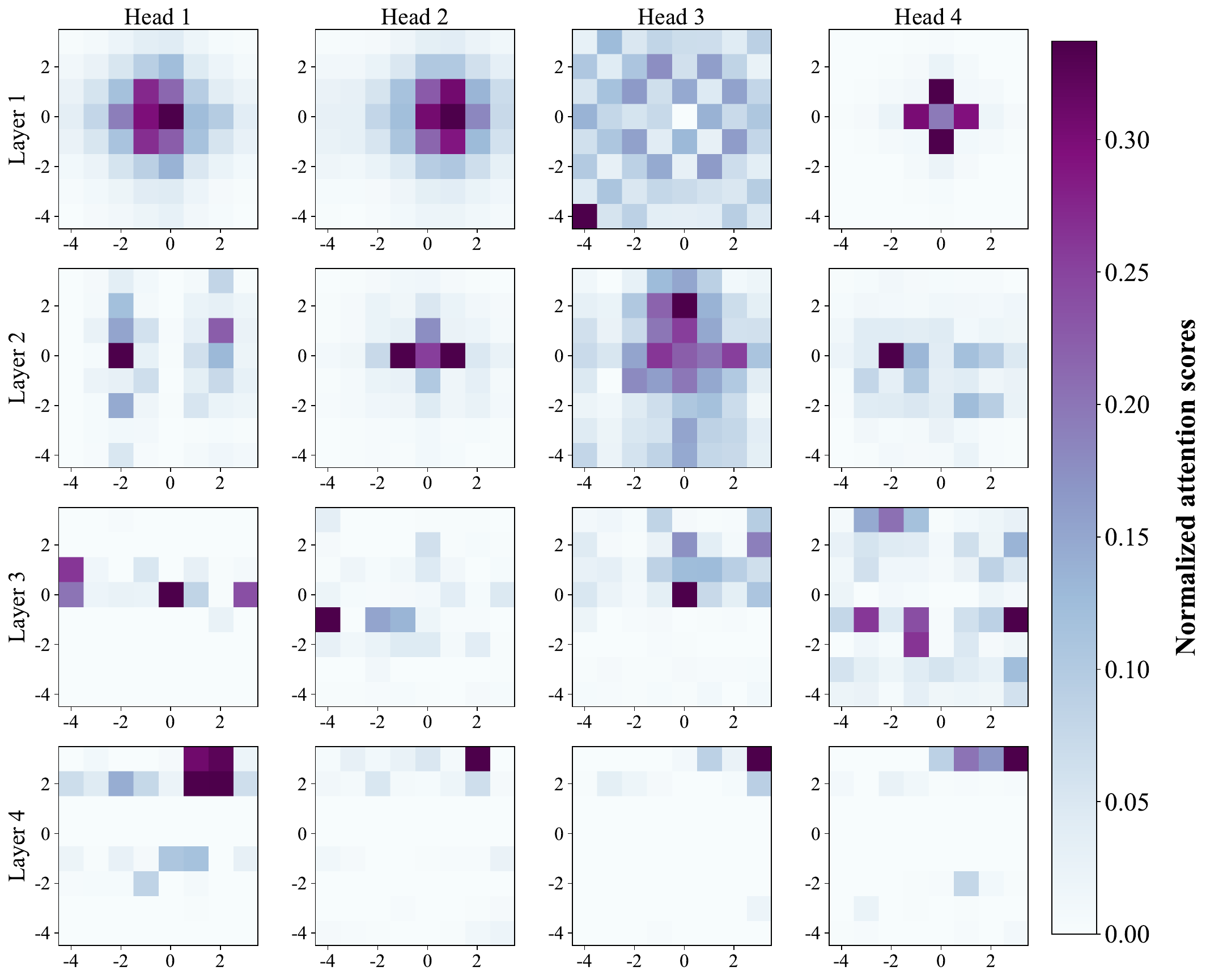}
    \caption{Similar as Figure~\ref{fig:score 8x8 half site} but for $8 \times 8$ Hubbard model with $t' = -0.2$ at $1/8$ dopping.}
    \label{fig:score 8x8 doped site}
\end{figure}

\begin{figure}
    \centering
    \includegraphics[width=\linewidth]{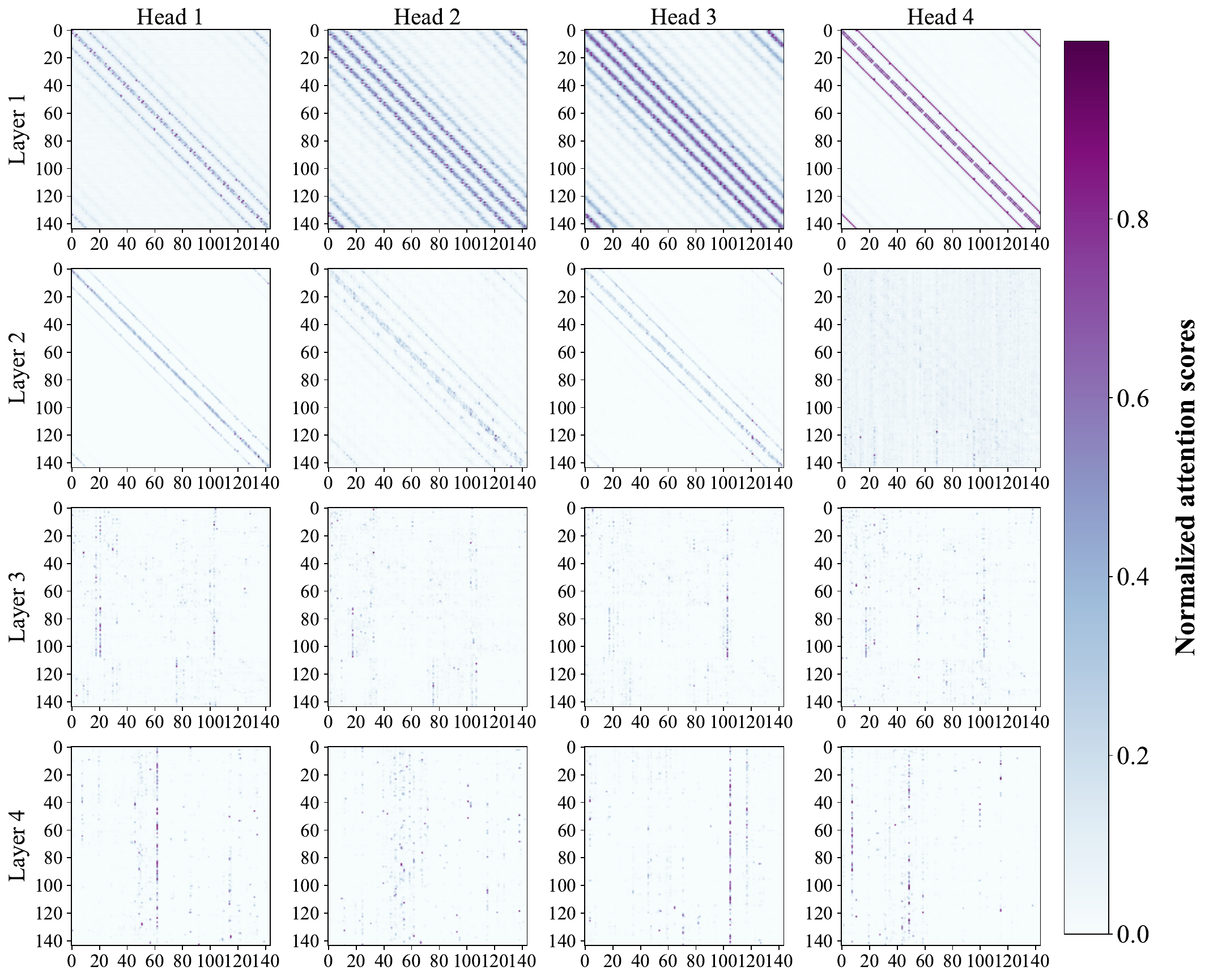}
    \caption{Similar as Figure~\ref{fig:score 8x8 half} but for $12 \times 12$ Hubbard model with $t' = -0.2$ at $1/8$ dopping.}
    \label{fig:score 12x12 doped}
\end{figure}

\begin{figure}
    \centering
    \includegraphics[width=\linewidth]{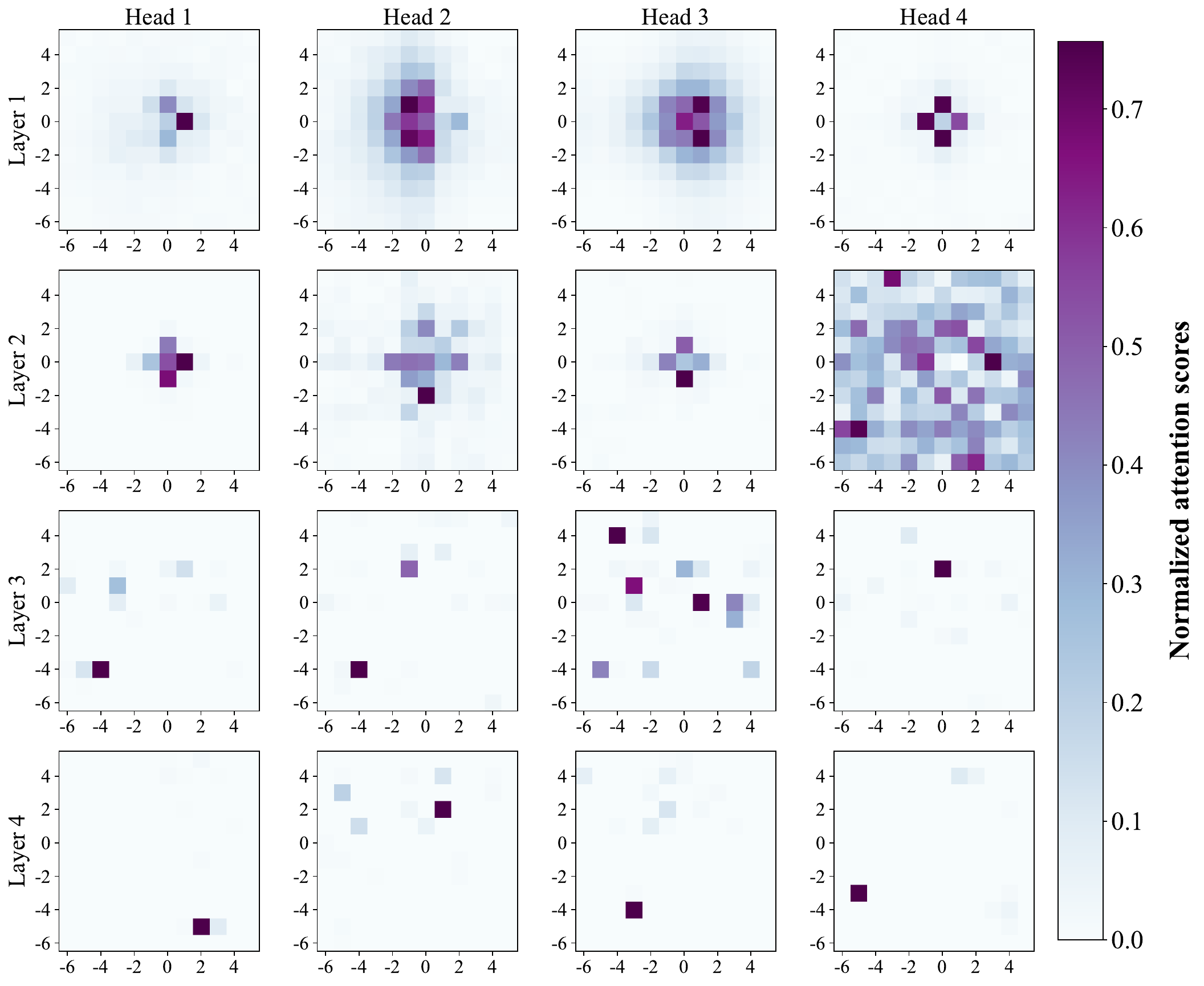}
    \caption{Similar as Figure~\ref{fig:score 8x8 half site} but for $12 \times 12$ Hubbard model with $t' = -0.2$ at $1/8$ dopping.}
    \label{fig:score 12x12 doped site}
\end{figure}

\begin{figure}
    \centering
    \includegraphics[width=0.59\linewidth]{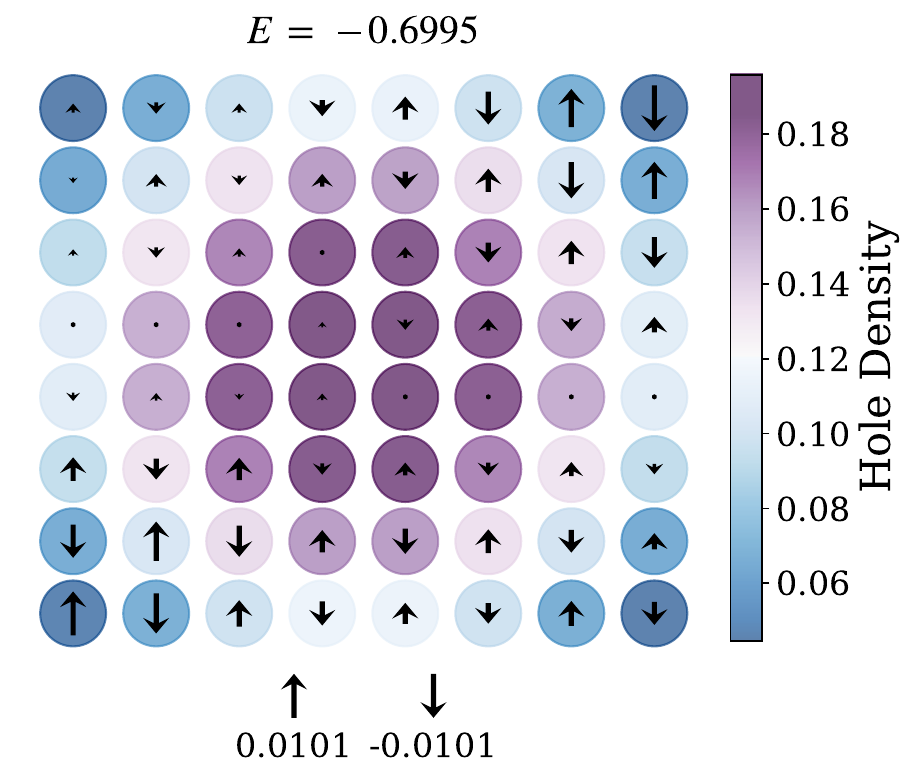}
    \caption{Hole and spin density distributions in the ground state for Hubbard with $U = 8$ and hole doping $\delta = 1/8$. The magnitudes of the spin density are represented by the sizes of the arrows while the direction is denoted by the direction of the arrows. Hole density is depicted using a color scale. The system is the pure Hubbard model with size $8 \times 8$ under OBC.}
    \label{fig:OBC 8x8}
\end{figure}

\begin{figure}
    \centering
    \includegraphics[width=0.85\linewidth]{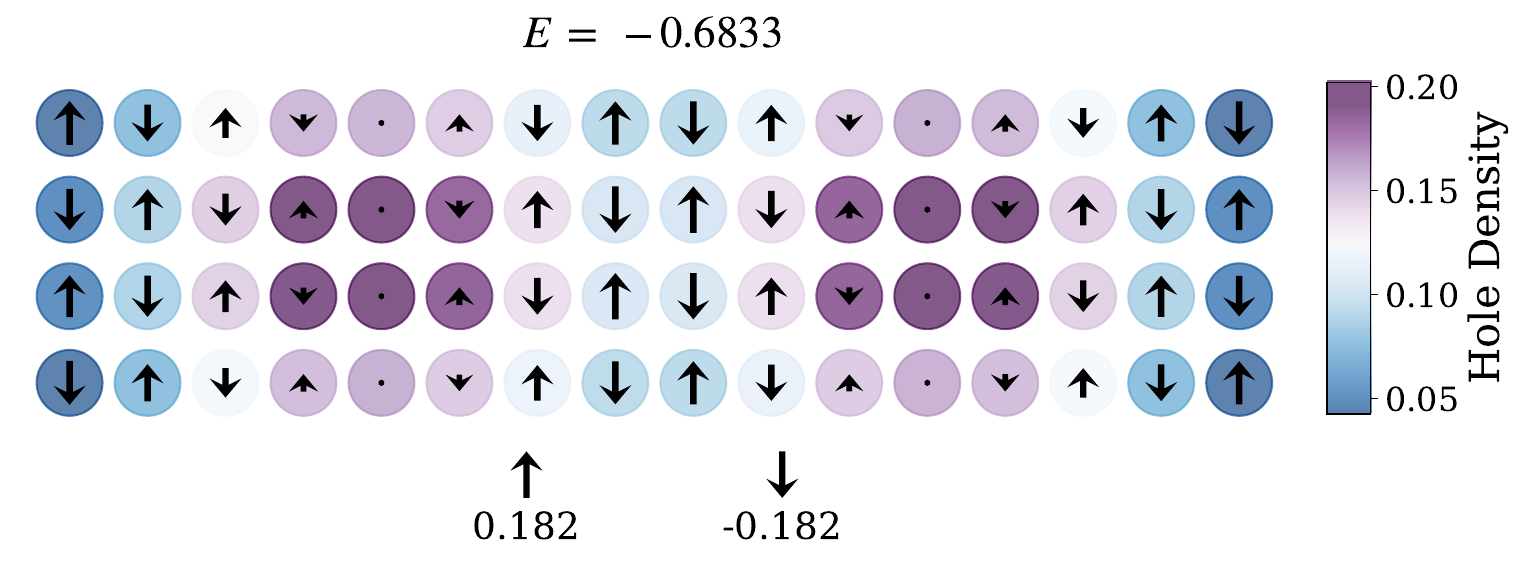}
    \caption{Similar as Figure~\ref{fig:OBC 8x8} but for $16 \times 4$ pure Hubbard model under OBC.}
    \label{fig:OBC 16x4}
\end{figure}

\begin{figure}
    \centering
    \includegraphics[width=0.9\linewidth]{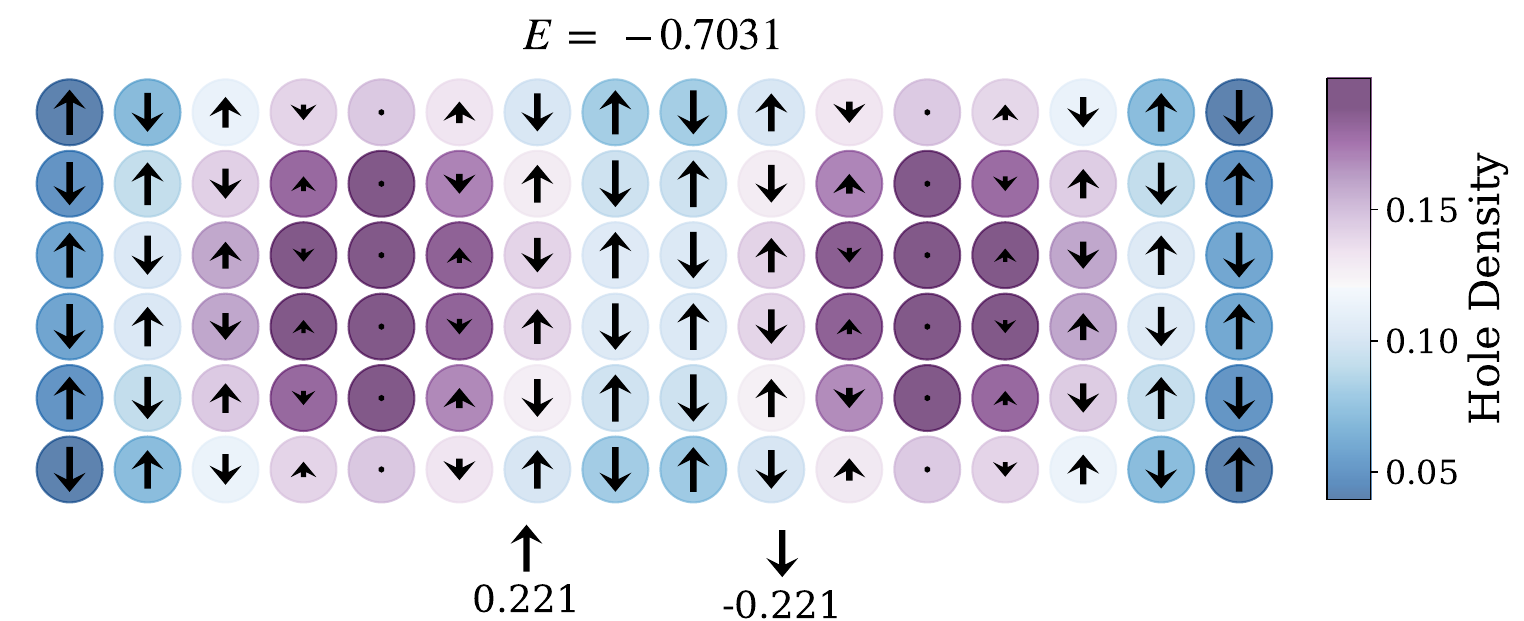}
    \caption{Similar as Figure~\ref{fig:OBC 8x8} but for $16 \times 6$ pure Hubbard model under OBC.}
    \label{fig:OBC 16x6}
\end{figure}

\begin{figure}
    \centering
    \includegraphics[width=0.88\linewidth]{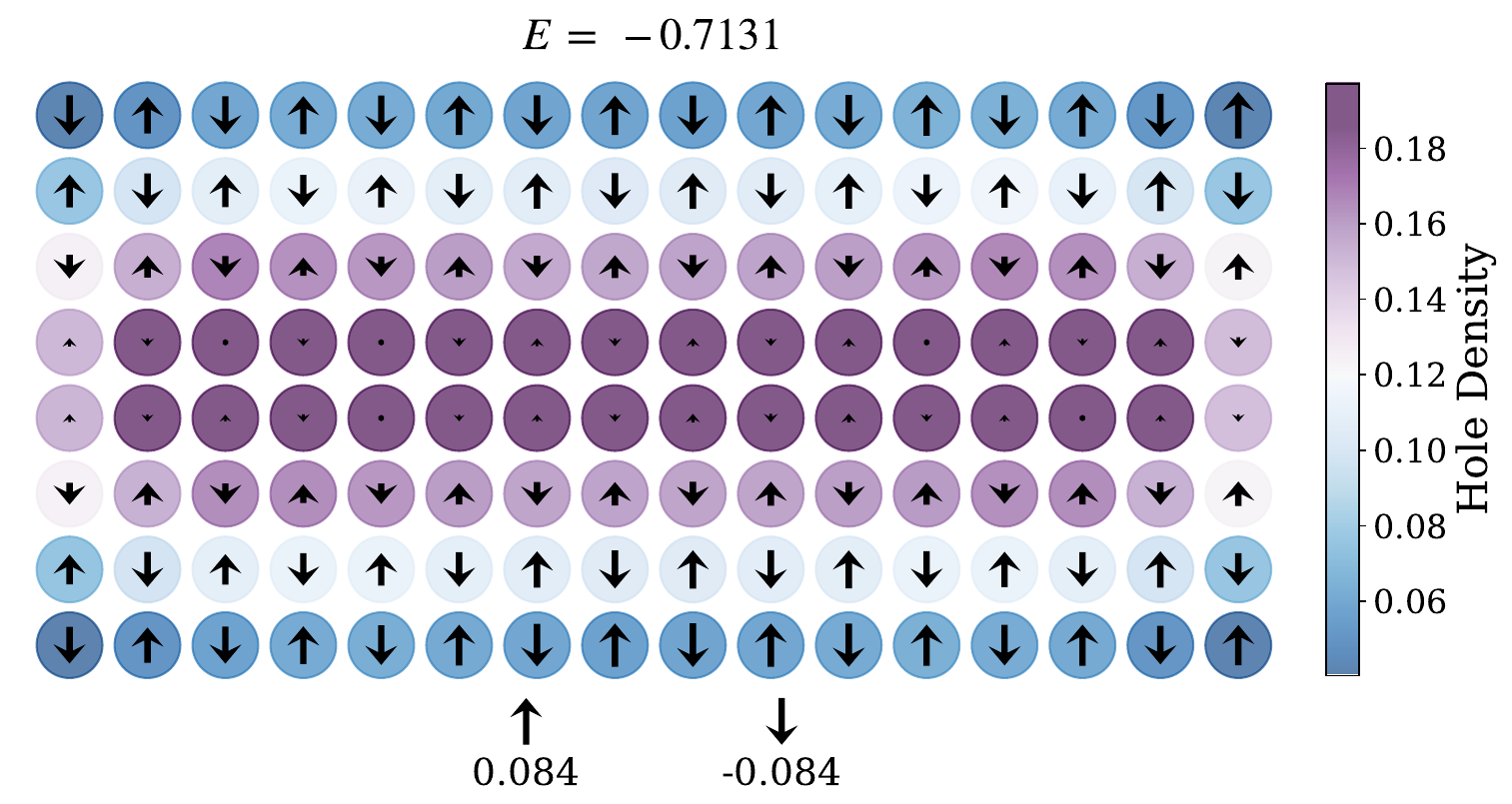}
    \caption{Similar as Figure~\ref{fig:OBC 8x8} but for $16 \times 8$ pure Hubbard model under OBC.}
    \label{fig:OBC 16x8}
\end{figure}

\begin{figure}
    \centering
    \includegraphics[width=0.9\linewidth]{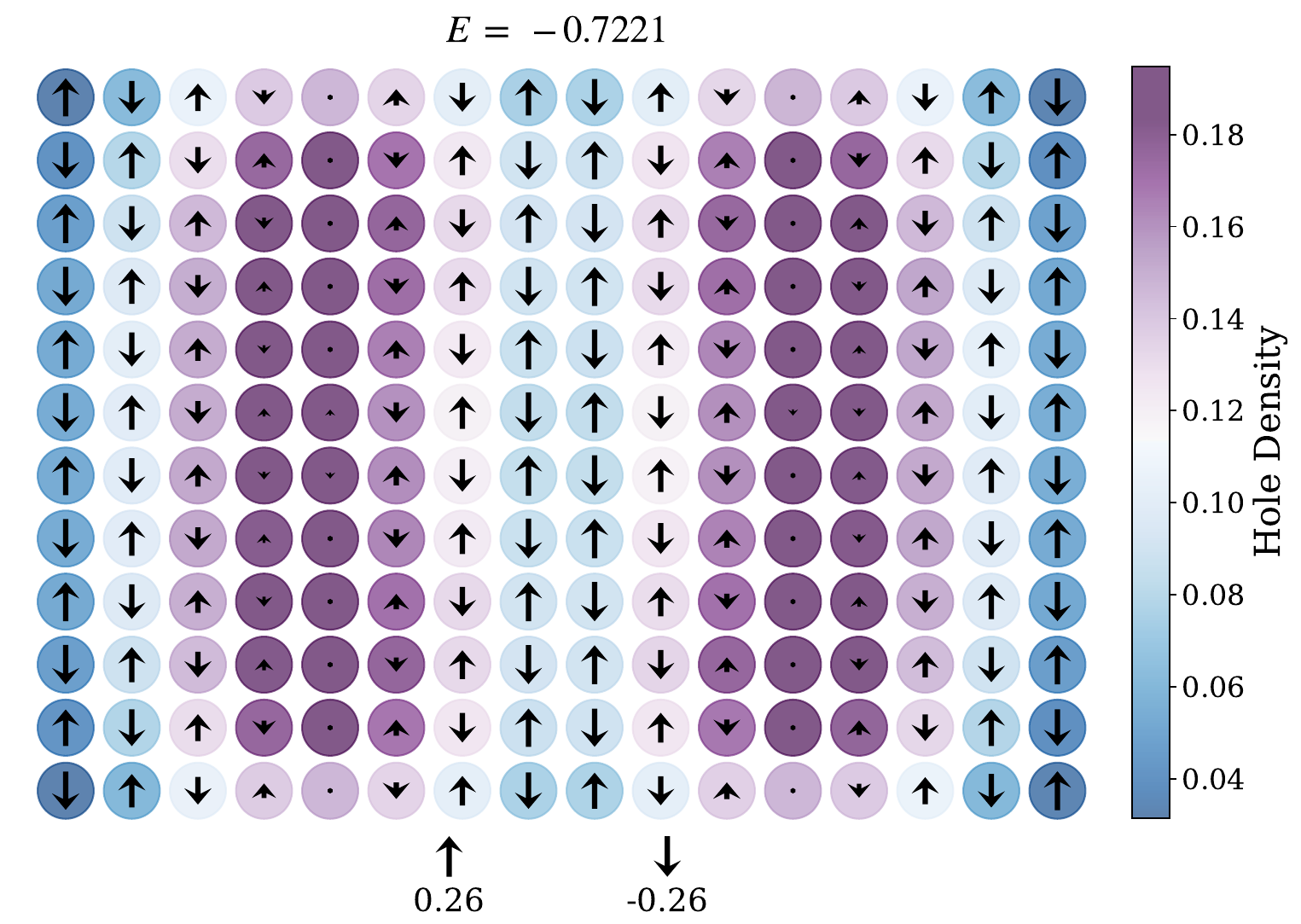}
    \caption{Similar as Figure~\ref{fig:OBC 8x8} but for $16 \times 12$ pure Hubbard model under OBC.}
    \label{fig:OBC 16x12}
\end{figure}

\begin{figure}
    \centering
    \includegraphics[width=0.9\linewidth]{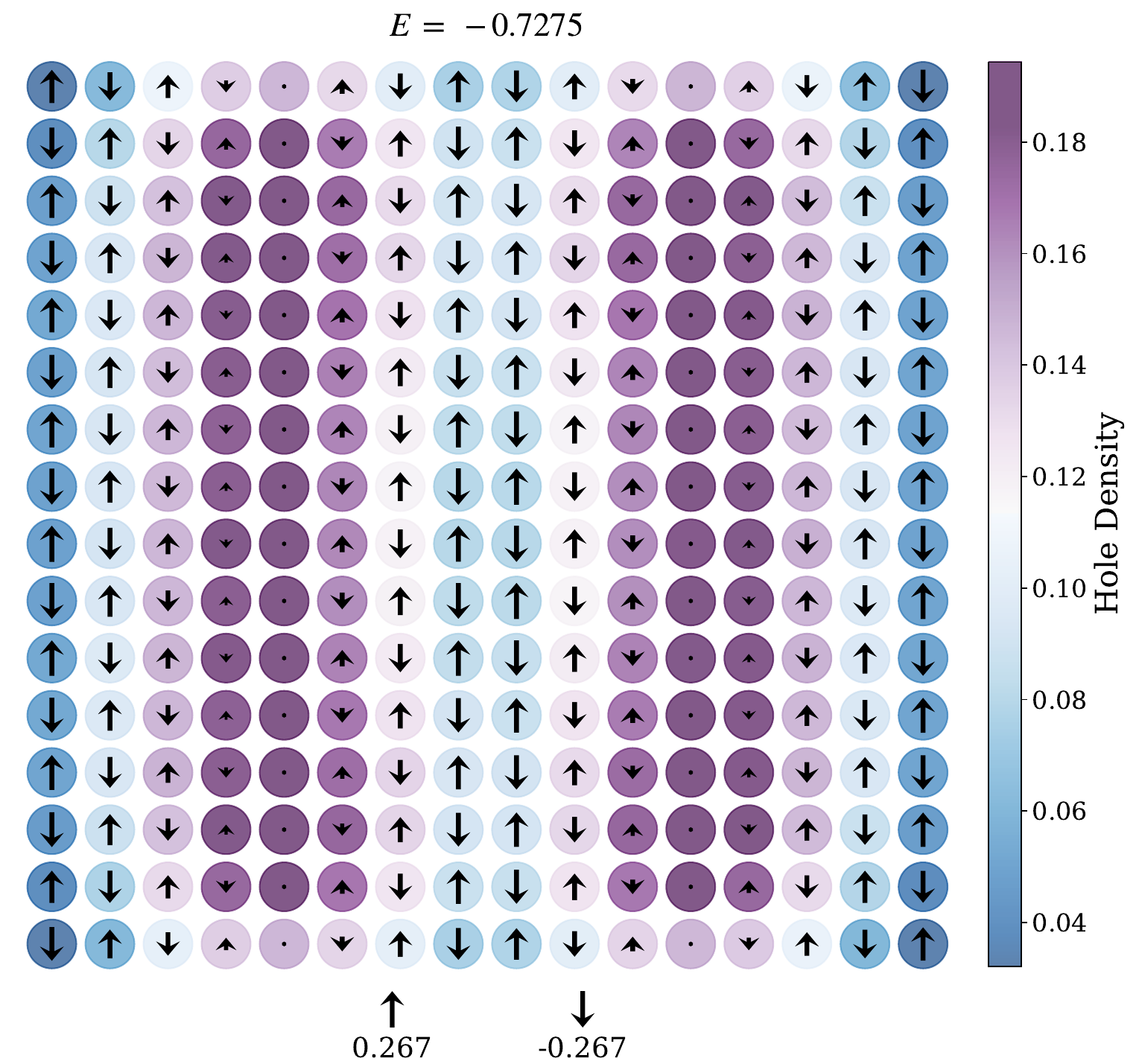}
    \caption{Similar as Figure~\ref{fig:OBC 8x8} but for $16 \times 16$ pure Hubbard model under OBC.}
    \label{fig:OBC 16x16}
\end{figure}

\begin{figure}
    \centering
    \includegraphics[width=0.85\linewidth]{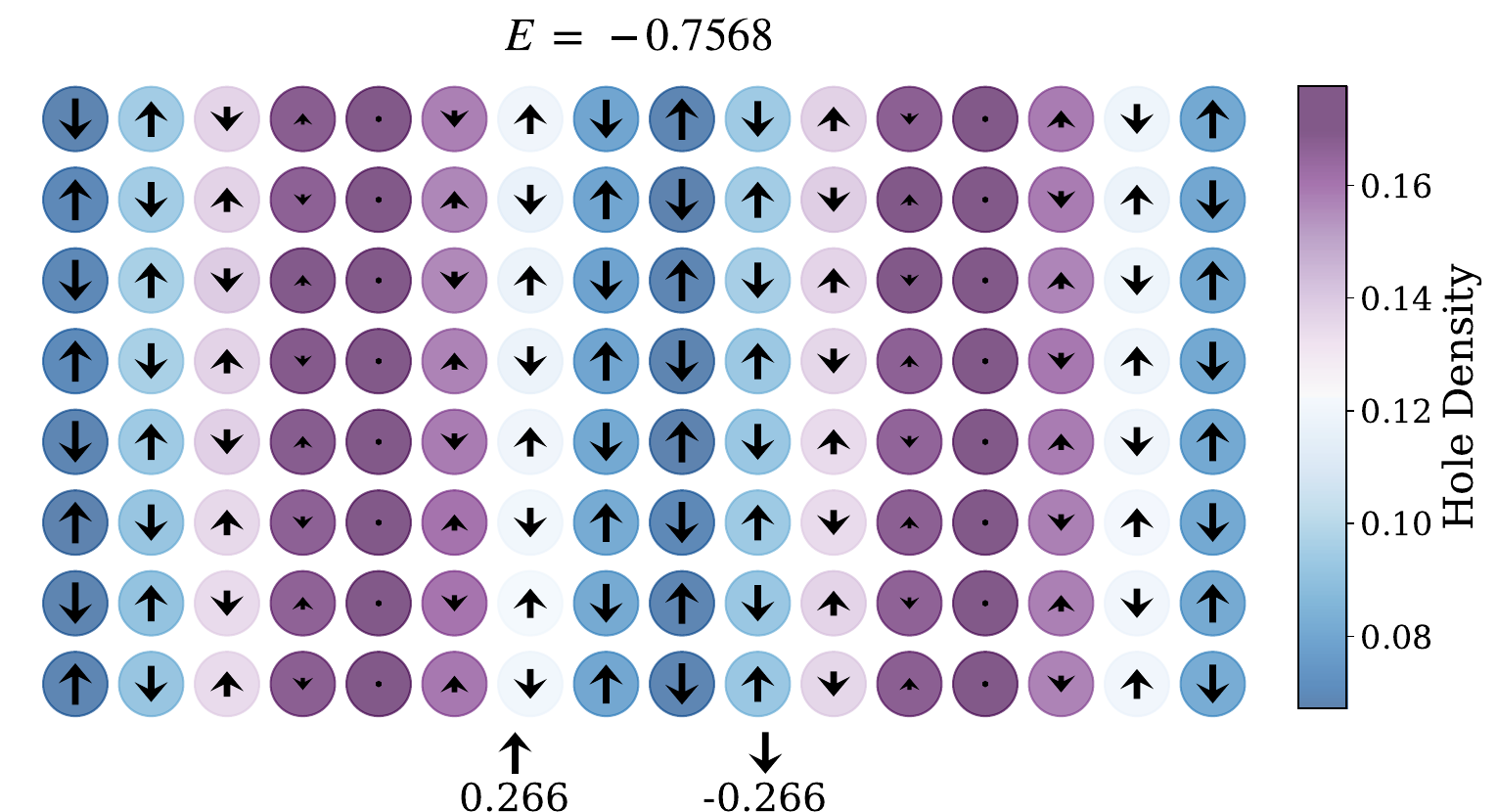}
    \caption{Similar as Figure~\ref{fig:OBC 8x8} but for $16 \times 8$ pure Hubbard model under PBC.}
    \label{fig:PBC 16x8}
\end{figure}

\begin{figure}
    \centering
    \includegraphics[width=0.88\linewidth]{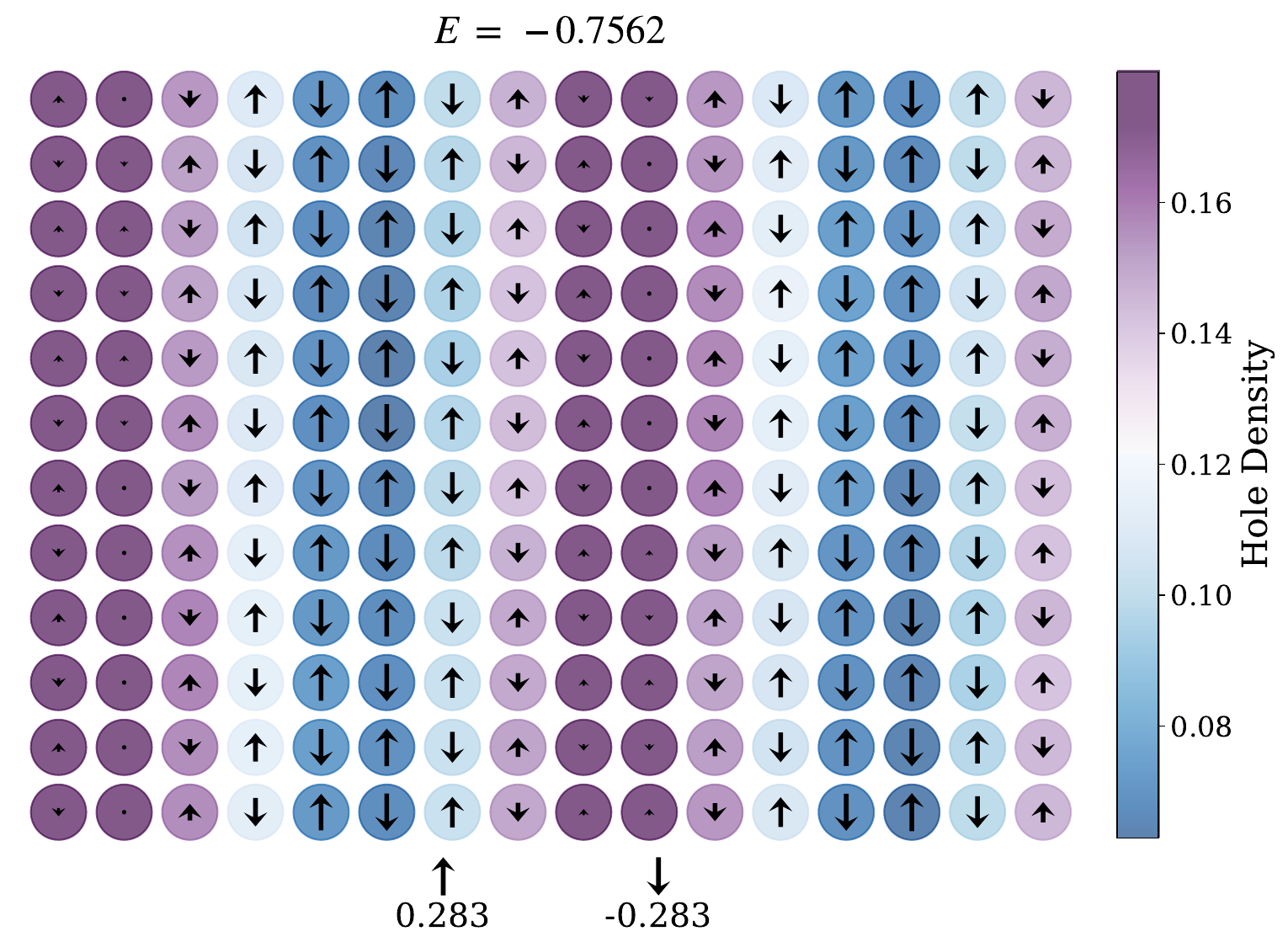}
    \caption{Similar as Figure~\ref{fig:OBC 8x8} but for $16 \times 12$ pure Hubbard model under PBC.}
    \label{fig:PBC 16x12}
\end{figure}

\begin{figure}
    \centering
    \includegraphics[width=0.55\linewidth]{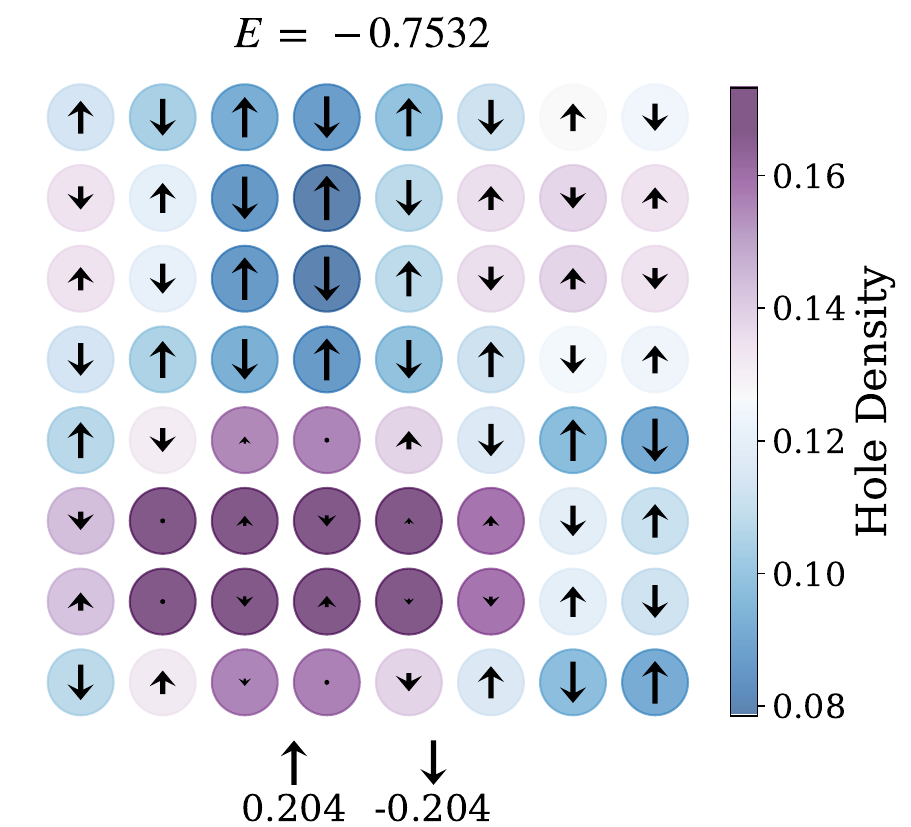}
    \caption{Similar as Figure~\ref{fig:OBC 8x8} but for $8 \times 8$ pure Hubbard model under PBC.}
    \label{fig:PBC 8x8}
\end{figure}

\begin{figure}
    \centering
    \includegraphics[width=0.77\linewidth]{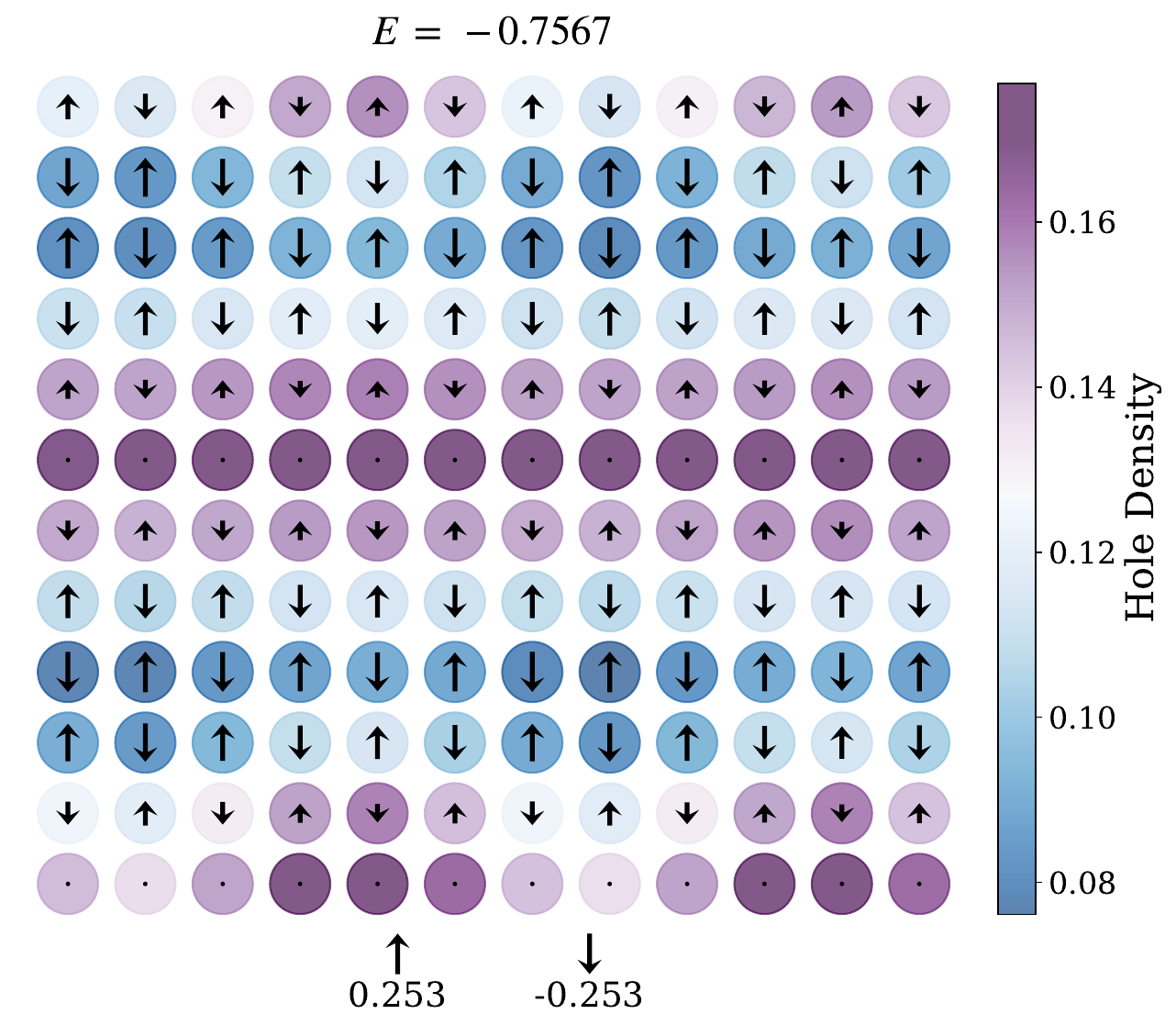}
    \caption{Similar as Figure~\ref{fig:OBC 8x8} but for $12 \times 12$ pure Hubbard model under PBC.}
    \label{fig:PBC 12x12}
\end{figure}

\begin{figure}
    \centering
    \includegraphics[width=0.54\linewidth]{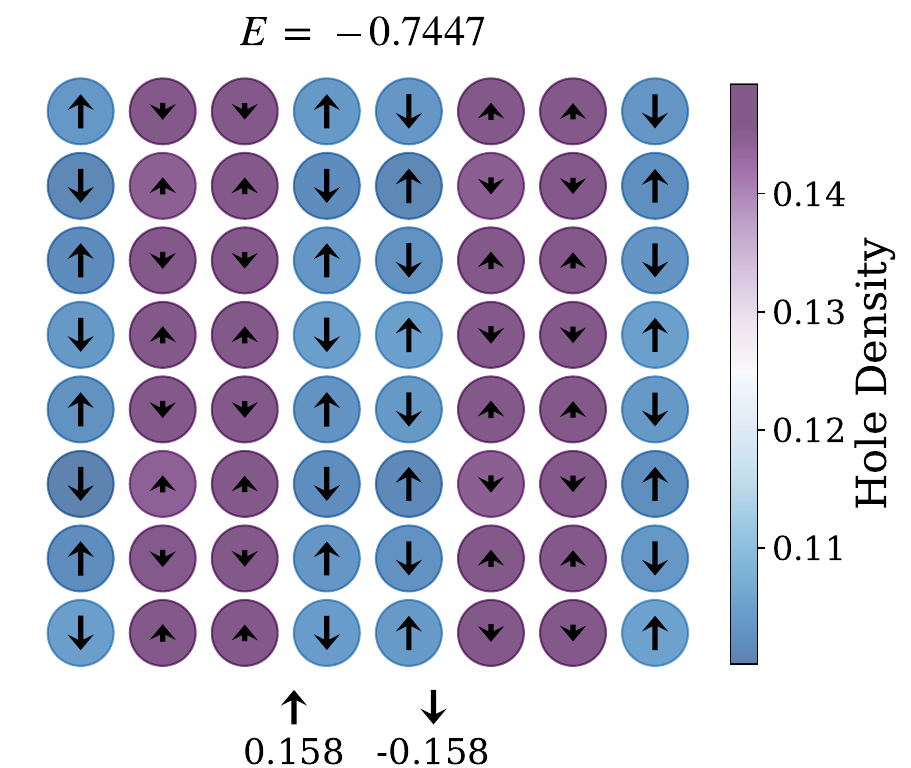}
    \caption{Similar as Figure~\ref{fig:OBC 8x8} but for $8 \times 8$ Hubbard model with $t'=-0.2$ under PBC.}
    \label{fig:PBC t2 8x8}
\end{figure}

\begin{figure}
    \centering
    \includegraphics[width=0.66\linewidth]{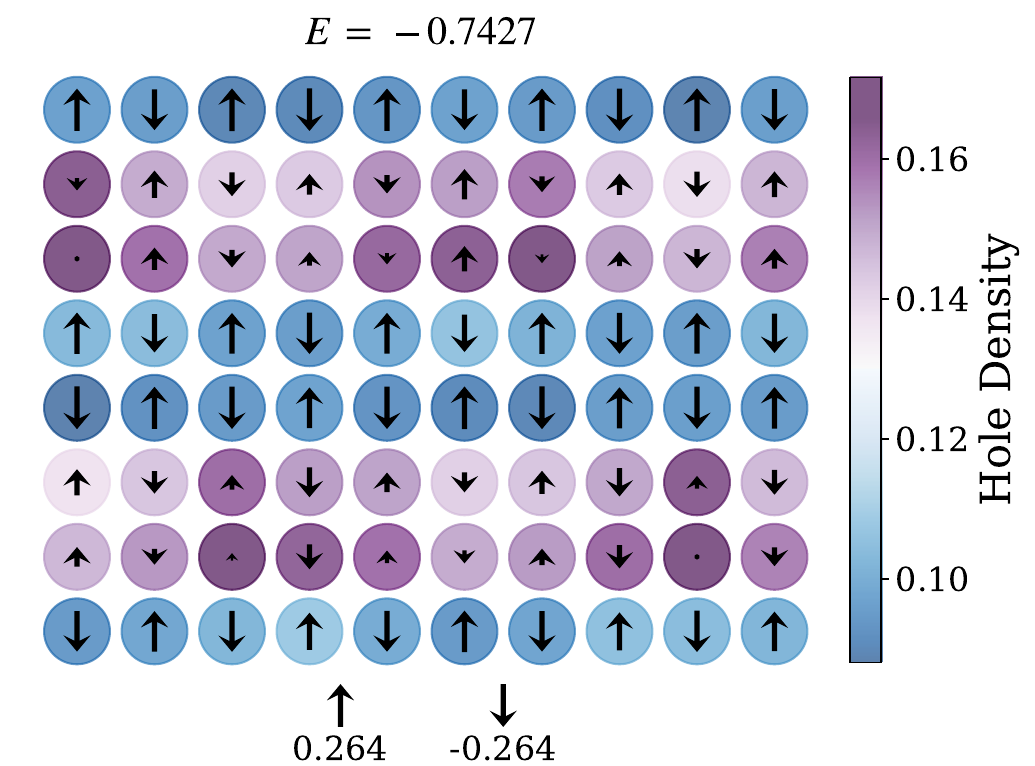}
    \caption{Similar as Figure~\ref{fig:OBC 8x8} but for $10 \times 8$ Hubbard model with $t'=-0.2$ under PBC.}
    \label{fig:PBC t2 10x8}
\end{figure}

\begin{figure}
    \centering
    \includegraphics[width=0.8\linewidth]{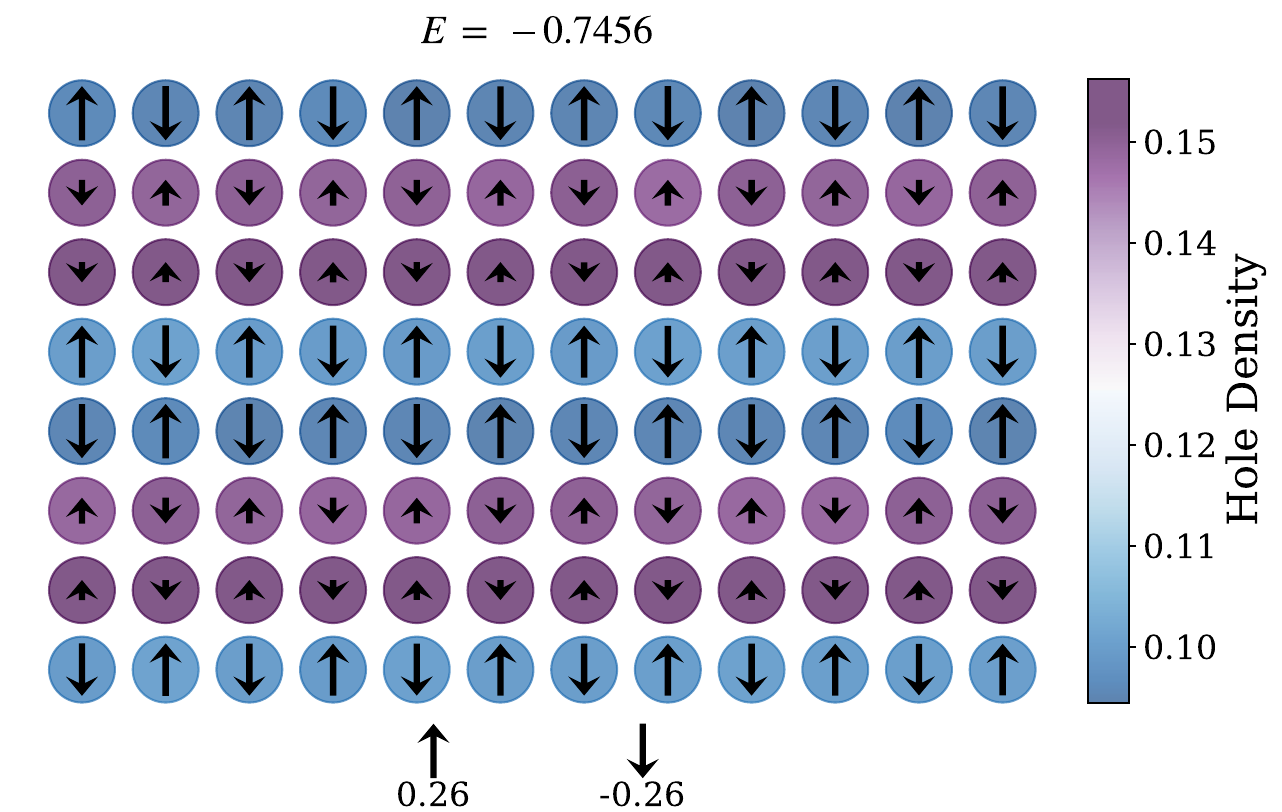}
    \caption{Similar as Figure~\ref{fig:PBC 16x16, PBC t2 16x12, 32x8 Combined} but for $12 \times 8$ Hubbard model with $t'=-0.2$ under PBC.}
    \label{fig:PBC t2 12x8}
\end{figure}

\begin{figure}
    \centering
    \includegraphics[width=0.93\linewidth]{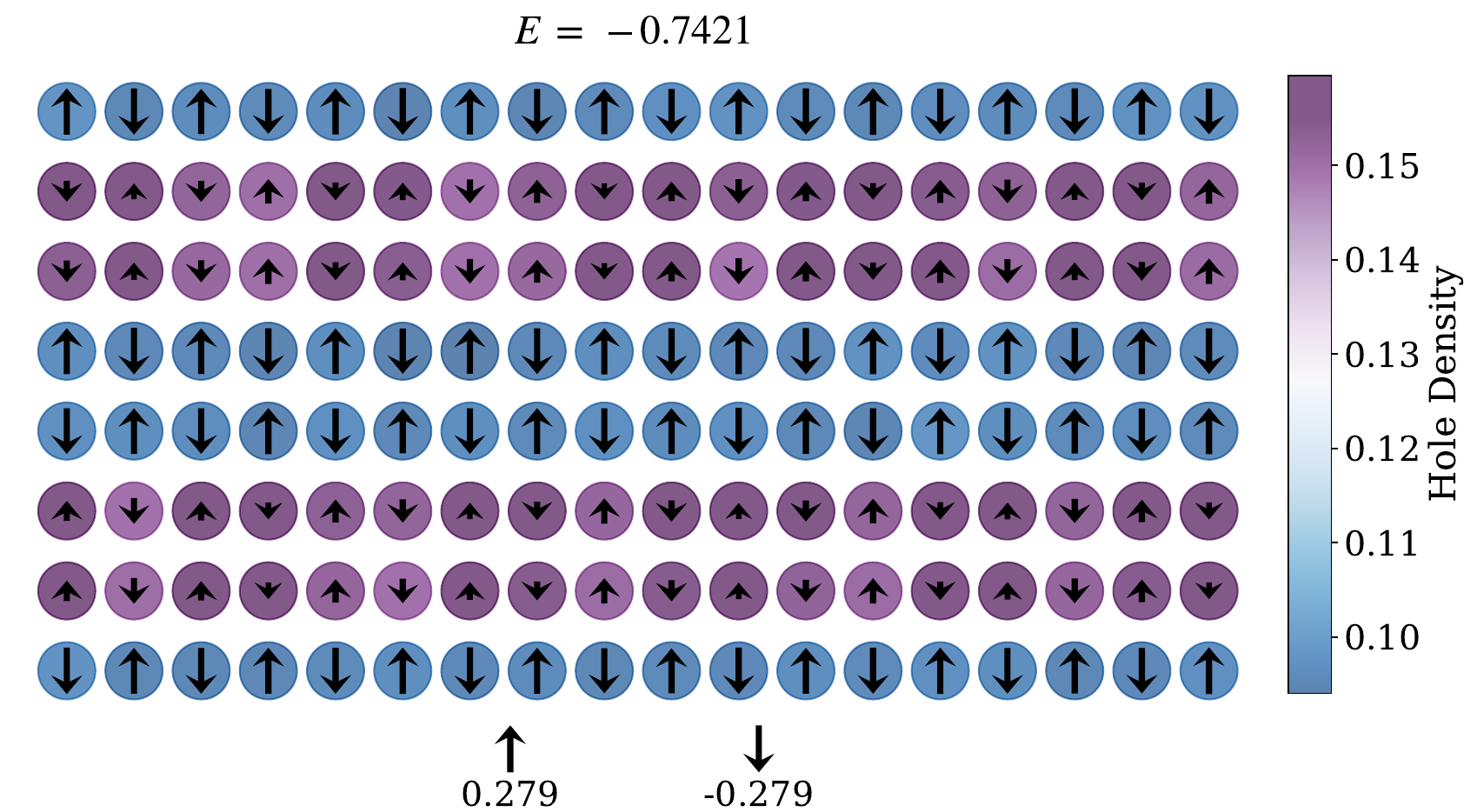}
    \caption{Similar as Figure~\ref{fig:OBC 8x8} but for $18 \times 8$ Hubbard model with $t'=-0.2$ under PBC.}
    \label{fig:PBC t2 18x8}
\end{figure}

\begin{figure}
    \centering
    \includegraphics[width=0.92\linewidth]{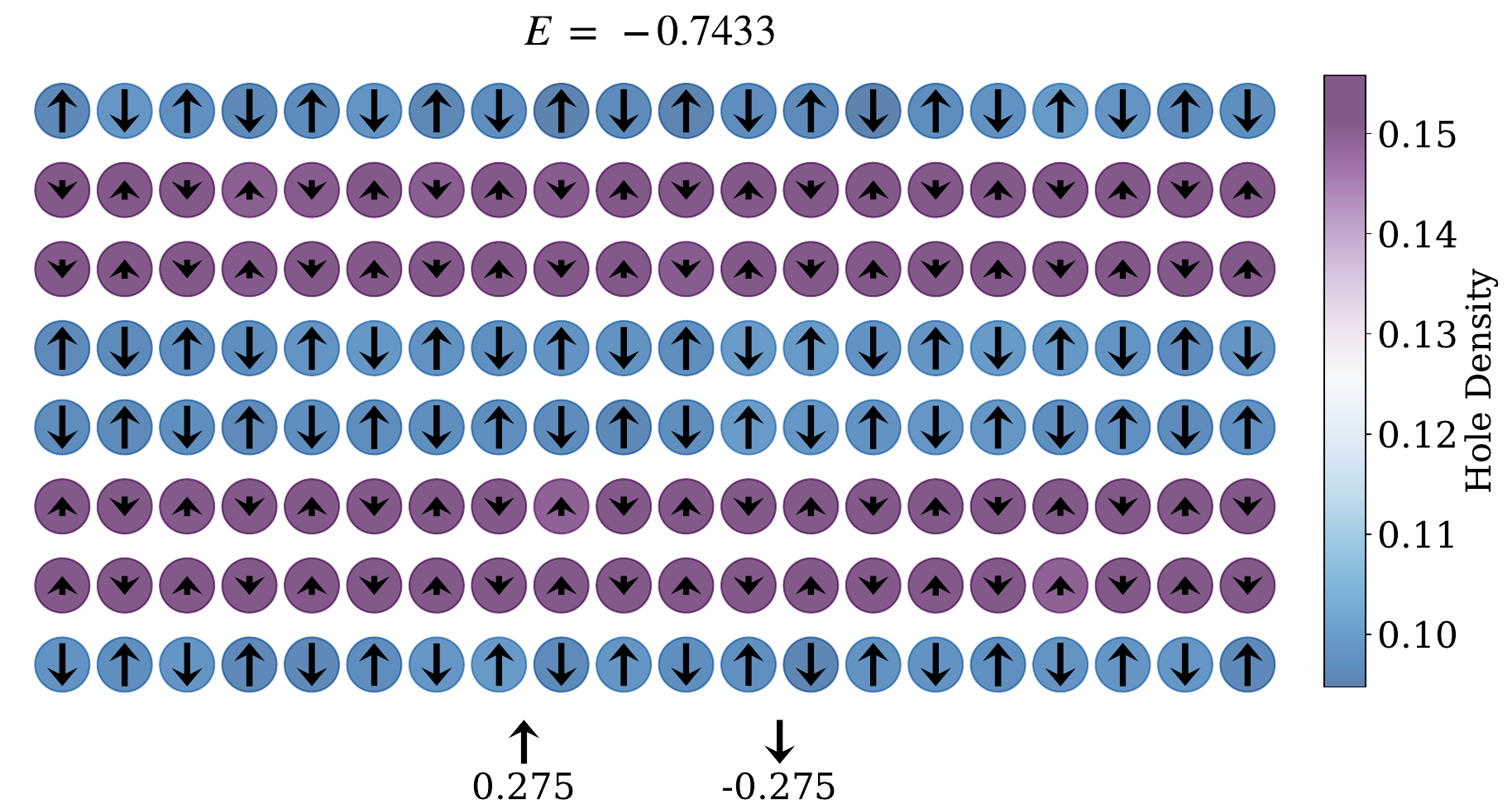}
    \caption{Similar as Figure~\ref{fig:OBC 8x8} but for $20 \times 8$ Hubbard model with $t'=-0.2$ under PBC.}
    \label{fig:PBC t2 20x8}
\end{figure}

\begin{figure}
    \centering
    \includegraphics[width=1\linewidth]{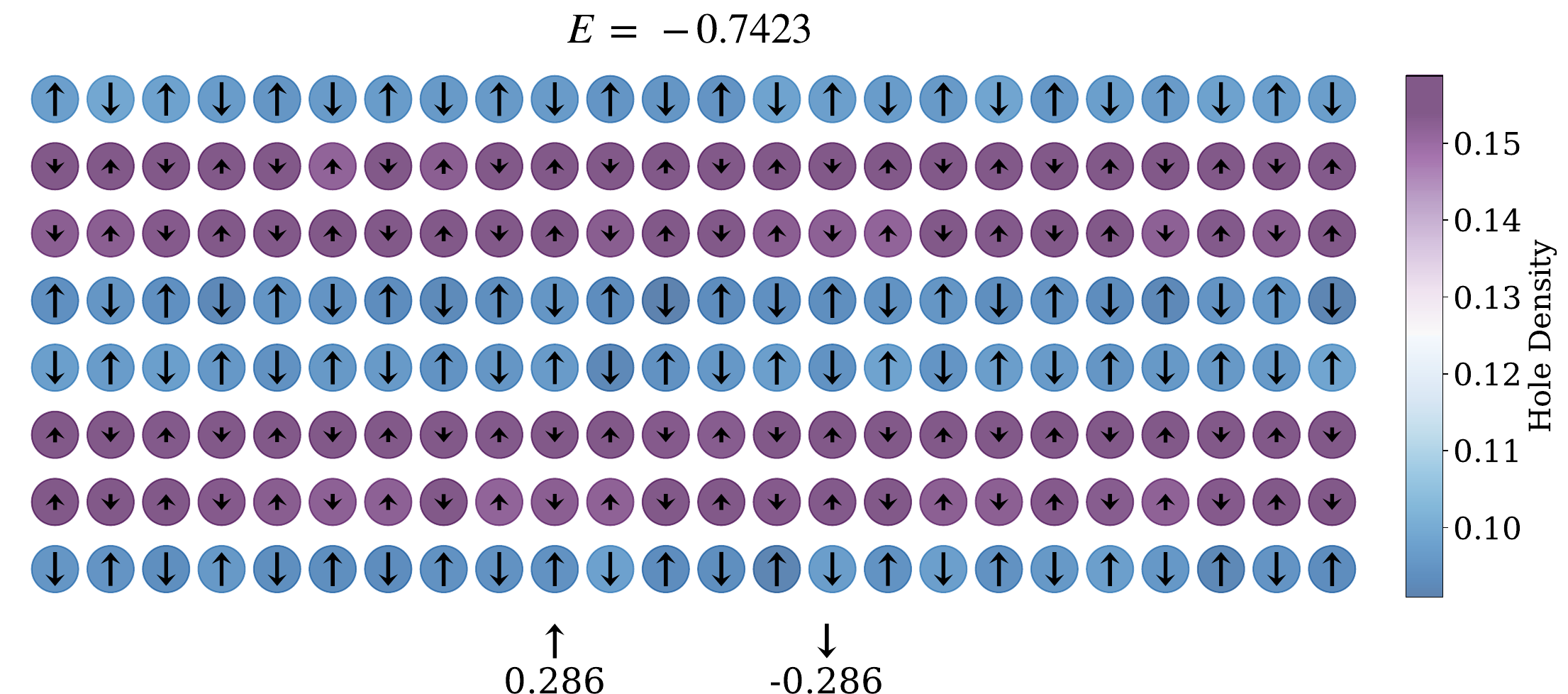}
    \caption{Similar as Figure~\ref{fig:OBC 8x8} but for $24 \times 8$ Hubbard model with $t'=-0.2$ under PBC.}
    \label{fig:PBC t2 24x8}
\end{figure}

\begin{figure}
    \centering
    \includegraphics[width=0.78\linewidth]{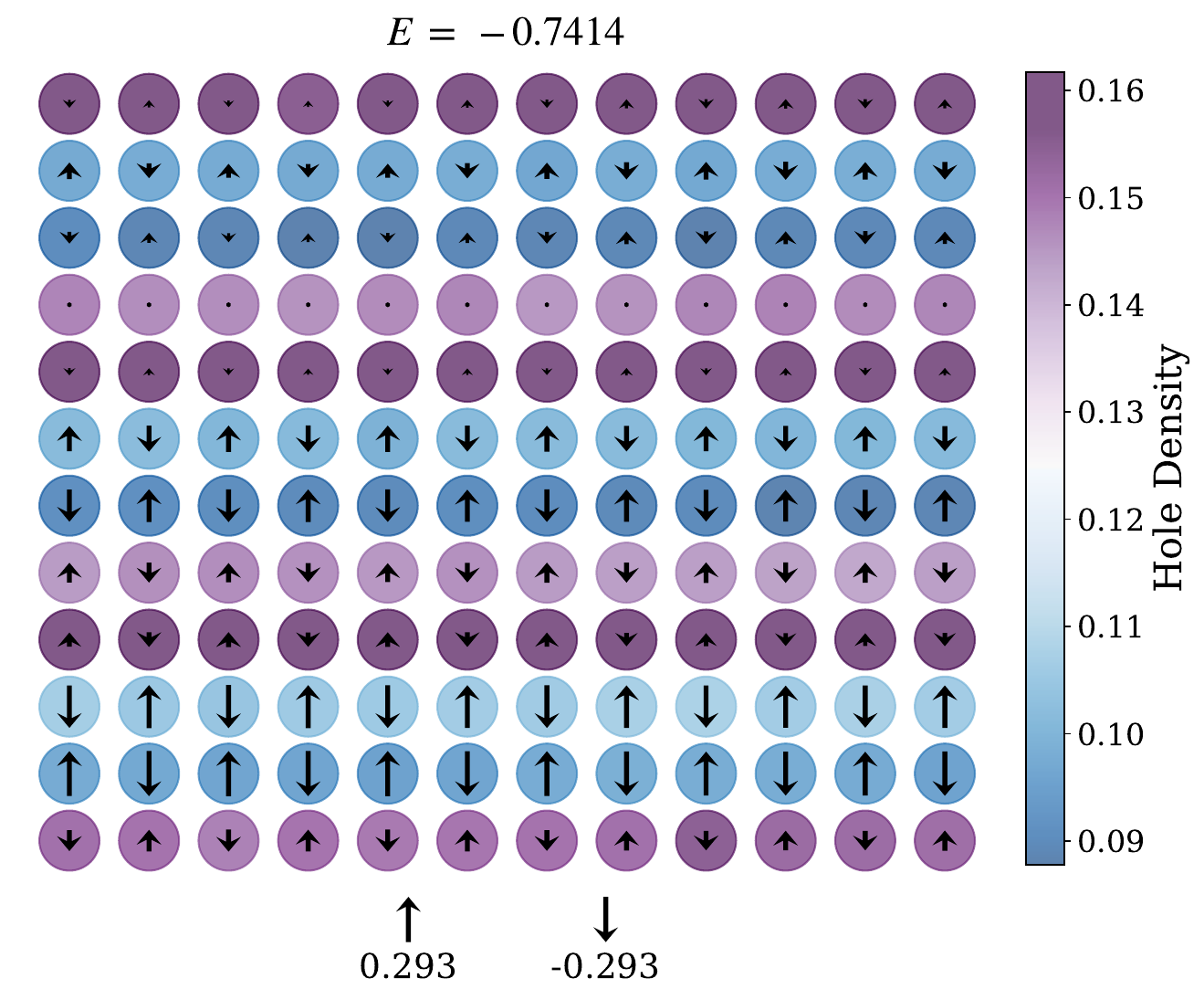}
    \caption{Similar as Figure~\ref{fig:OBC 8x8} but for $12 \times 12$ Hubbard model with $t'=-0.2$ under PBC.}
    \label{fig:PBC t2 12x12}
\end{figure}

\begin{figure}
    \centering
    \includegraphics[width=0.93\linewidth]{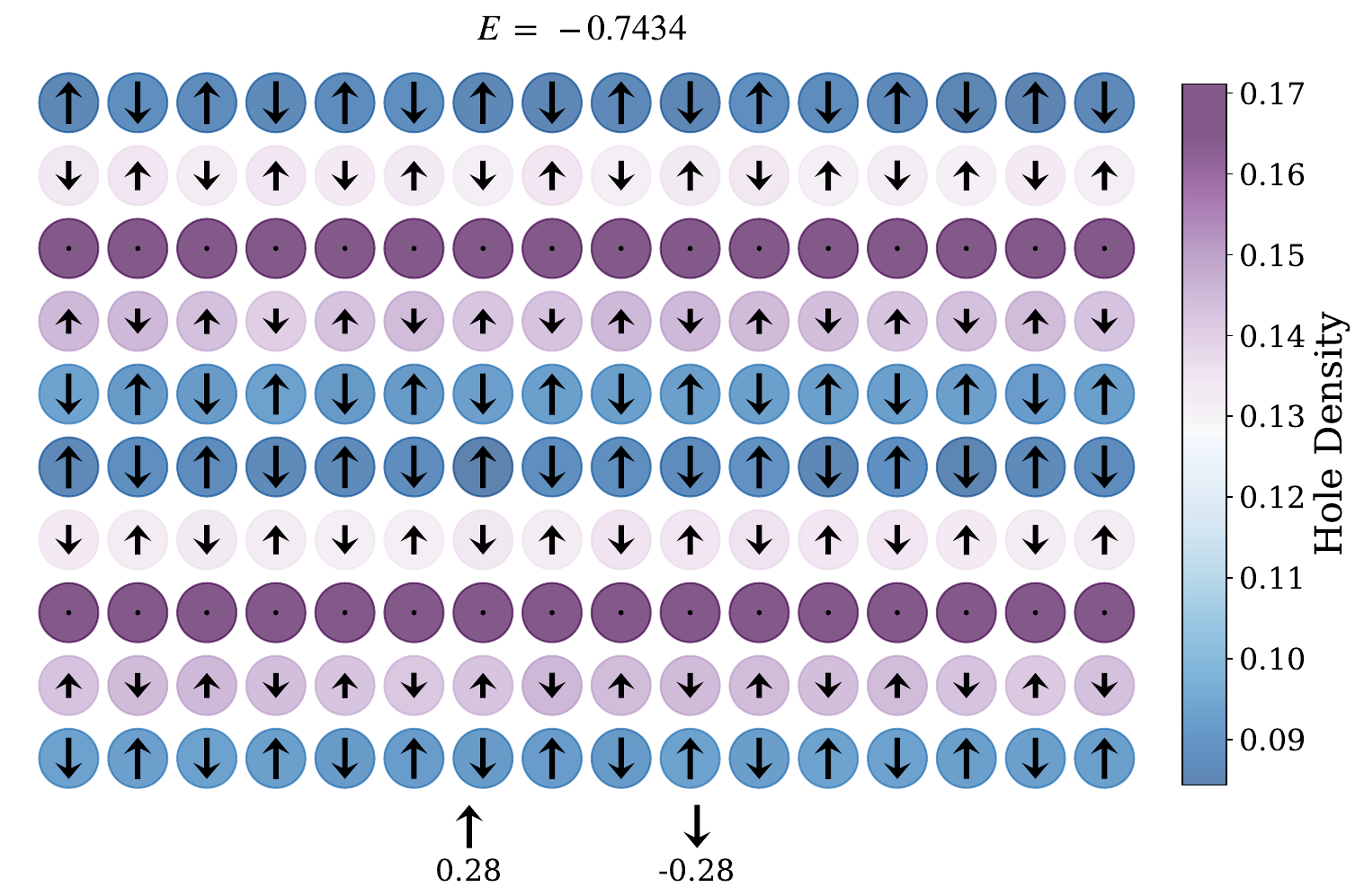}
    \caption{Similar as Figure~\ref{fig:OBC 8x8} but for $16 \times 10$ Hubbard model with $t'=-0.2$ under PBC.}
    \label{fig:PBC t2 16x10}
\end{figure}

\begin{figure}
    \centering
    \includegraphics[width=0.9\linewidth]{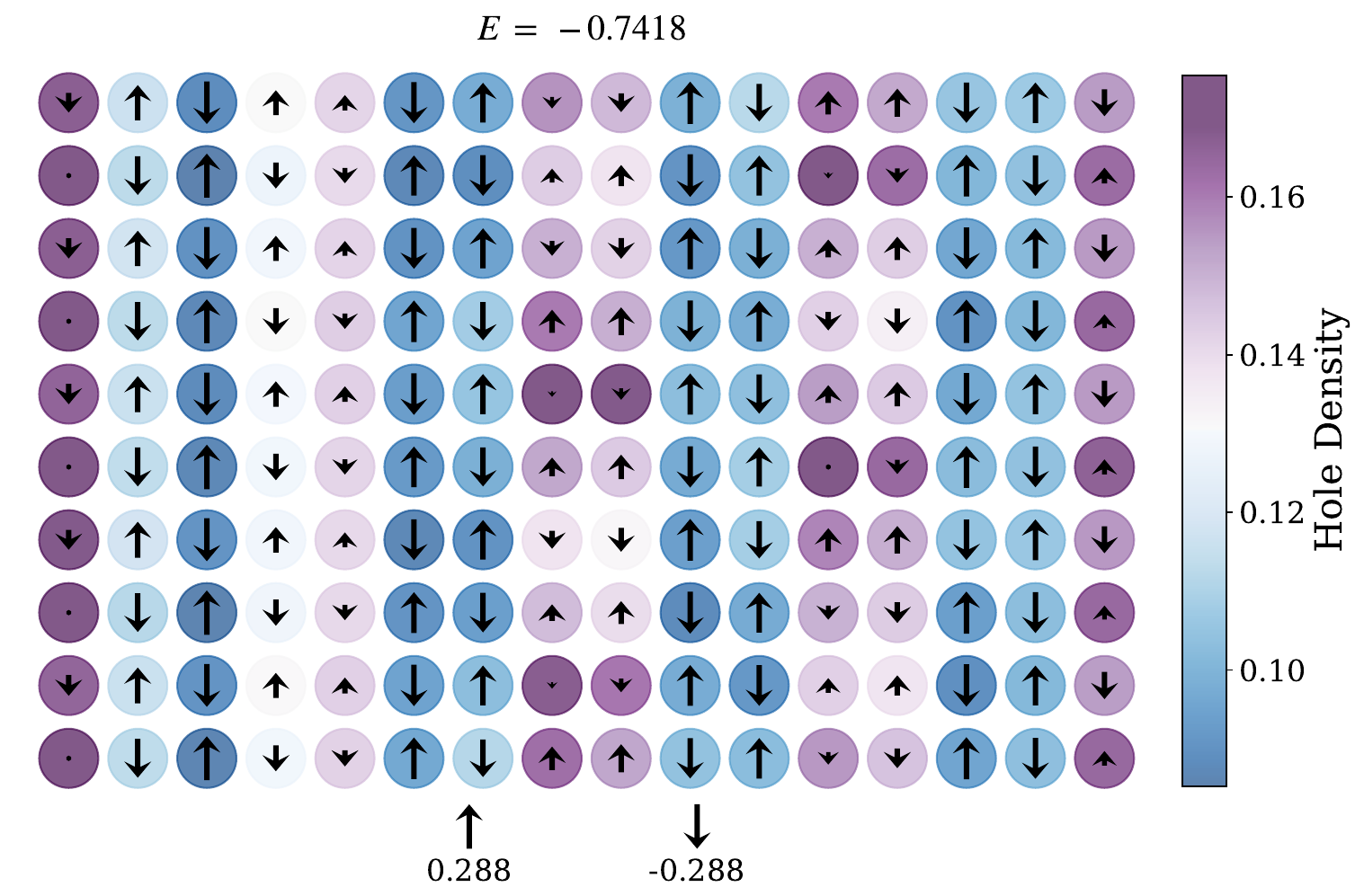}
    \caption{Similar as Figure~\ref{fig:OBC 8x8} but for $16 \times 10$ Hubbard model with $t'=-0.2$ under PBC. We apply a temporary pinning field to get the vertical stripe.}
    \label{fig:PBC t2 16x10 new}
\end{figure}

\begin{figure}
    \centering
    \includegraphics[width=0.87\linewidth]{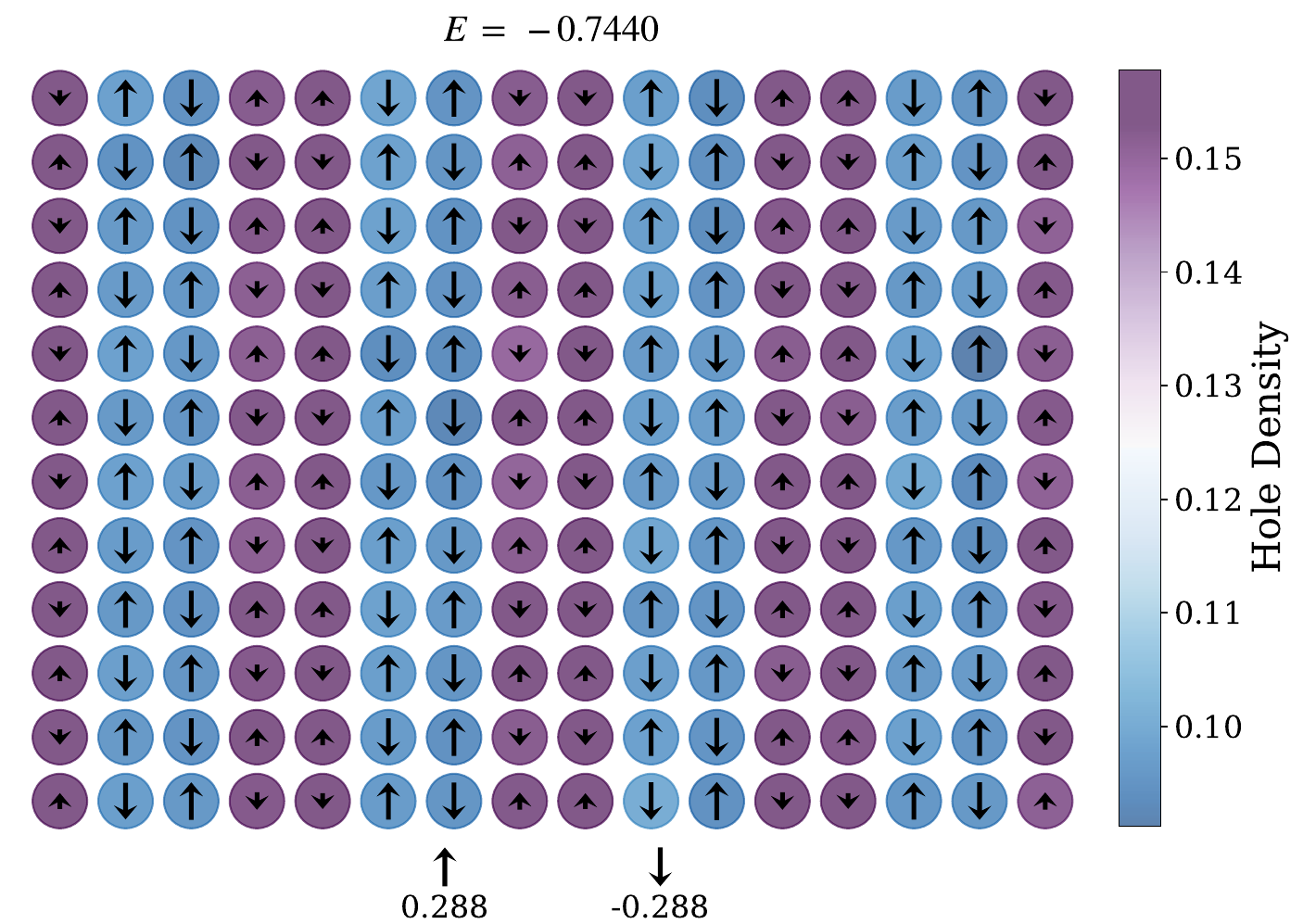}
    \caption{Similar as Figure~\ref{fig:OBC 8x8} but for $16 \times 12$ Hubbard model with $t'=-0.2$ under PBC.}
    \label{fig:PBC t2 16x12}
\end{figure}

\begin{figure}
\centering
\includegraphics[width=1\textwidth]{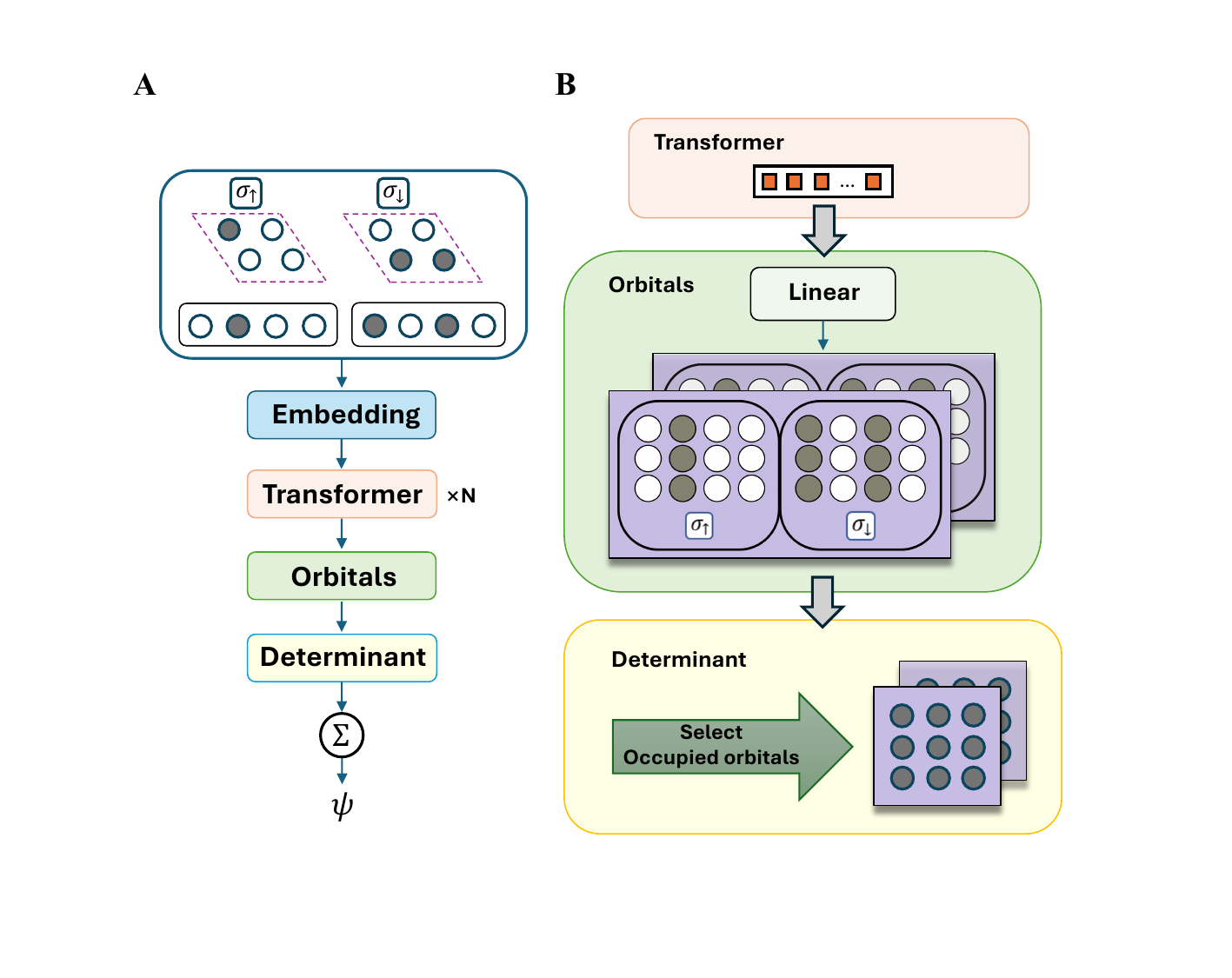}
\caption{\textbf{Overview of our NQS wavefunction architecture.} (a) This figure gives an example of a $2\times2$ Hubbard model. The color of the sites denotes the occupations of electrons on the sites. The process of embedding produces the same number of tokens as the number of sites, which are given as input to the neural network.
(b) Details of each building block. The transformer will output $K$ sets of backflow orbitals, which are selected according to the occupations of electrons. The final wavefunction amplitude is the sum of all determinants.}
\label{fig:arc}
\end{figure}

\begin{figure} %
	\centering
	\includegraphics[width=0.65\textwidth]{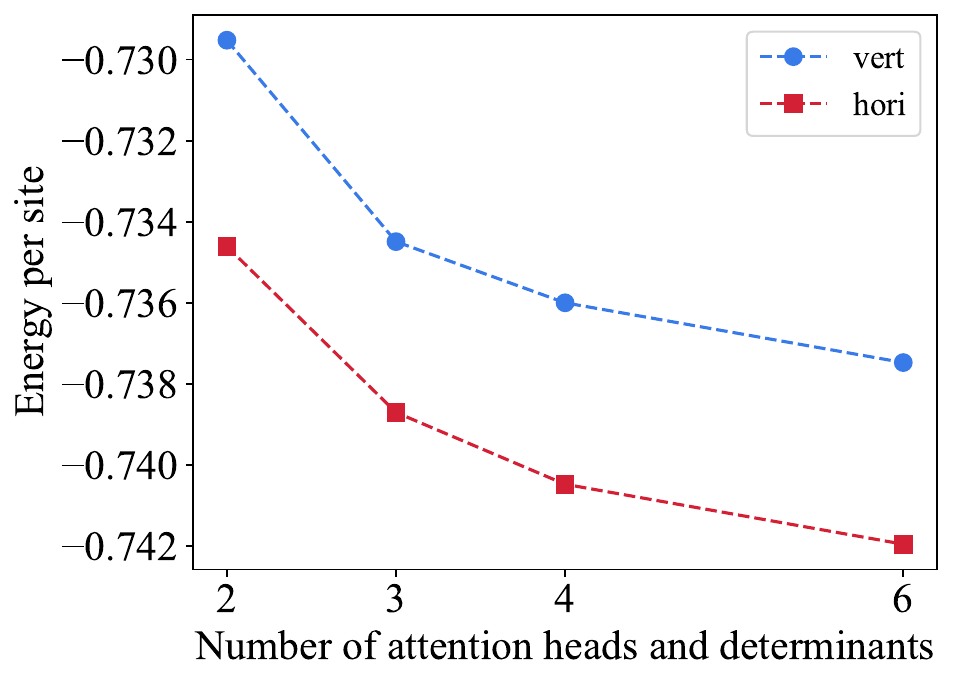} 
	\caption{Vertical v.s. Horizontal stripe for the $32 \times 8$ Hubbard model with $t'=-0.2$ and $\delta = 1/8$, using NQS with up to $6$ attention heads and determinants.}
	\label{fig:size_ablation} %
\end{figure}

\begin{figure} %
	\centering
	\includegraphics[width=0.95\textwidth]{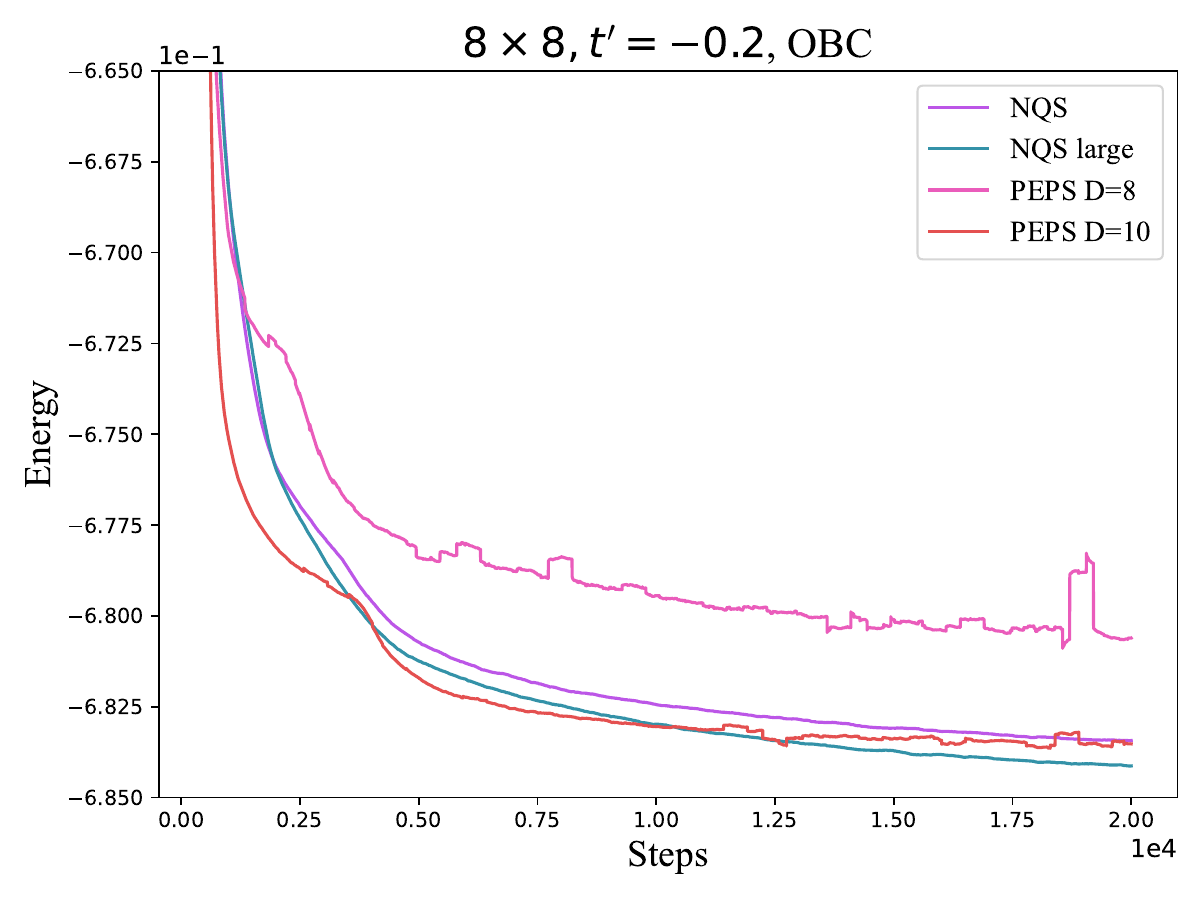} %

	\caption{\textbf{NQS v.s. PEPS under the same VMC batchsize}. NQS denotes the hidden dimension of 256 and 4 determinants, NQS large denotes the hidden dimension of 384 and 6 determinants. D denotes the bond dimension in PEPS. The expressive power of NQS is similar to PEPS with $D=10$.}
	\label{fig:gpu_peps} %
\end{figure}

\begin{figure} %
	\centering
	\includegraphics[width=0.95\textwidth]{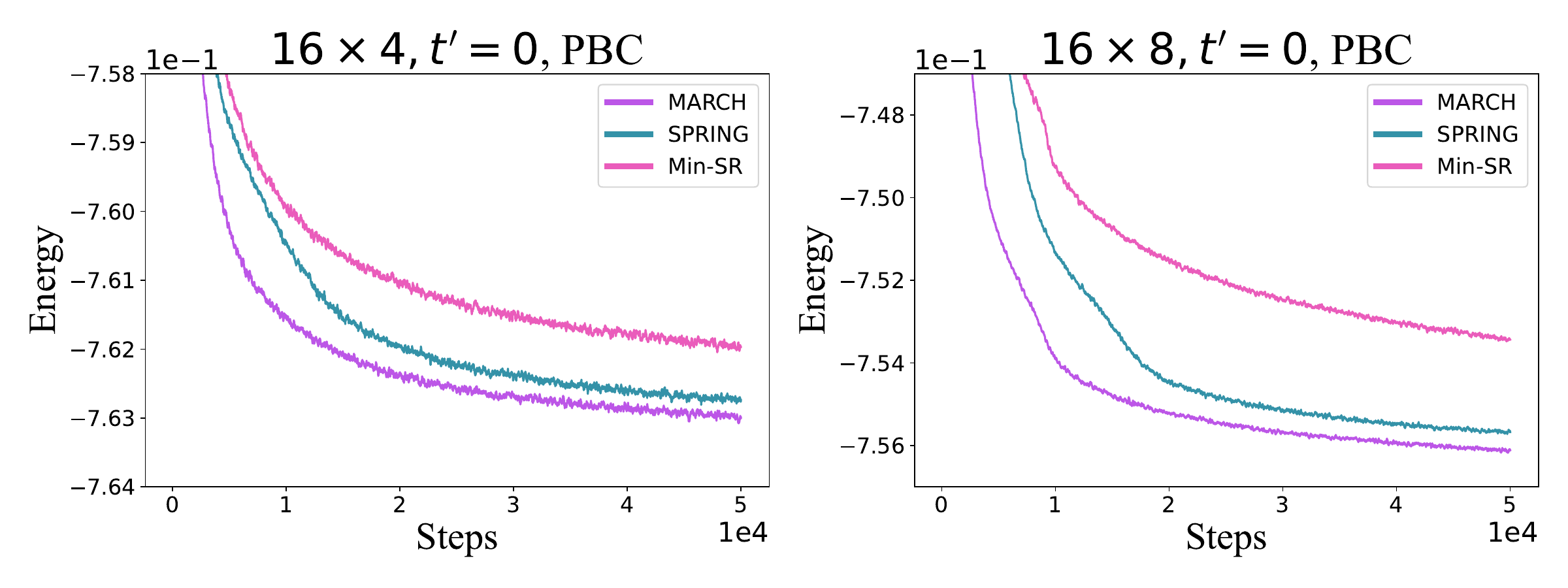} %

	\caption{\textbf{Ablation study of optimizer}. The results show that our optimizer converges significantly faster than SPRING and Min-SR.}
	\label{fig:optimizer_ablation} %
\end{figure}

\begin{table} %
\centering
\caption{\textbf{Benchmark energy in pure Hubbard model at half-filling with PBC.}
NQS achieve comparable results with the numerically exact AFQMC values \cite{qin2016benchmark}.}
\label{tab:afqmc} %

\begin{tabular}{c|ccc} %
    \hline
    Systems & $8\times 8$ & $10\times 10$ & $12\times 12$\\
    \hline
    AFQMC & -0.5262(5) & -0.5254(3) & -0.5246(1) \\
    NQS & -0.52582 & -0.52492 & -0.52440 \\
    \hline
\end{tabular}
\end{table}

\begin{table} %
	\centering
	\caption{\textbf{Benchmark energy in pure Hubbard model with OBC.}
		NQS achieve new state-of-the-art results by surpassing PEPS.}
	\label{tab:main} %

	\begin{tabular}{c|cccccc} %
		\hline
		Systems & $16\times 4$ & $16\times 6$ & $8\times 8$ & $16\times 8$ & $16\times 12$ & $16\times 16$\\
		\hline
		DMRG & -0.68537 & -0.70550 & -0.69840 & -0.71250 & \\
		PEPS & -0.68304(5) & -0.7008(2) & -0.69928(3) & -0.7122(3) & -0.7202(2) & -0.7260(2)\\
		NQS & -0.68325 & -0.70307 & -0.69952 & -0.71309 & -0.72212 & -0.72747\\
            Hartree-Fock & -0.52499 & -0.54269 &  -0.53874 & -0.55335 & -0.56250 & -0.56716\\
		\hline
	\end{tabular}
\end{table}

\begin{table} %
	\centering
	\caption{\textbf{Computation complexity relative to NQS under the same hardware condition in $8\times 8$ lattice}. NQS is significantly more efficient than PEPS.}
	\label{tab:speed} %

	\begin{tabular}{c|cccc} %
		\hline
		Methods & NQS & NQS large & PEPS D=8 & PEPS D=10\\
		\hline
		Wall Time & $1\times$ & $2\times$ & $7\times$ & $24\times$ \\
		\hline
	\end{tabular}
\end{table}

\begin{table} %
	\centering
	\caption{\textbf{Accuracy on simple systems, where the exact solution is possible}.}
	\label{tab:estimate} %

	\begin{tabular}{c|cccccc} %
		\hline
		Algorithm & \cite{nomura2017restricted} & \cite{luo2019backflow} & \cite{inui2021determinant} & \cite{robledo2022fermionic} & \cite{zhou2024solving} & This work\\
		\hline
            Lattice & $8\times 8$ & $4\times 4$ & $4\times 4$ & $4\times 4$ & $8\times 8$ & $12\times 12$\\
            \hline
            Dopping & $0$ & $1/8$ & $3/8$ & $1/8$ & $0$ & $0$\\
            \hline
		Relative Error & $0.003$ & $0.01$ & $0.0005$ & $0.001$ & $0.005$ & $0.0002$ \\
		\hline
            Accuracy & $99.7\%$ & $99\%$ & $99.95\%$ & $99.9\%$ & $99.5\%$ & $99.98\%$ \\
            \hline
	\end{tabular}
\end{table}

\begin{table}[h]
  \centering
  \begin{tabular}{l|l}
    \toprule
    \midrule
     Config & Value \\
    \midrule
    Optimizer & Adam \\
    Optimizer momentum & $\mu$, $\beta$= 0.9, 0.999\\
    Batch size & 4096 \\
    Learning rate at time $t$ & $10 ^{-4}(1 + t/10^4)^{-1}$\\
    Local energy clipping & 5.0 \\
    MCMC step & $2.5L_x\times Ly$\\
    Hidden dimension & 256\\
    Layers & 2\\
    Steps & 20000\\
    \bottomrule
  \end{tabular}
  \caption{List of configs for the NNB training.}
  \label{tab:nnb}
\end{table}

\begin{table}[h]
  \centering
  \begin{tabular}{l|l}
    \toprule
    \midrule
     Config & Value \\
    \midrule
    Optimizer & Adam \\
    Batch size & 4096 \\
    Learning rate & $3 \times 10 ^{-4}$ \\
    MCMC step & $30$\\
    Steps & 5000\\
    \bottomrule
  \end{tabular}
  \caption{List of configs for the pretraining phase.}
  \label{tab:pretrain}
\end{table}

\begin{table}[h]
  \centering
  \begin{tabular}{l|l}
    \toprule
    \midrule
     Config & Value \\
    \midrule
    Optimizer & MARCH \\
    Optimizer momentum & $\mu$, $\beta$= 0.95, 0.995\\
    Damping & $\lambda = 0.001$\\
    Batch size & 4096 \\
    Norm constraint at time $t$ & $10 ^{-1}(1 + \text{max}(t - 8000, 0)/8000)^{-1}$\\
    Local energy clipping & 5.0 \\
    MCMC step & $2.5L_x\times Ly$\\
    Hidden dimension & 256\\
    Layers & 4\\
    Number of determinants & 4 \\
    Steps & 100000\\
    \bottomrule
  \end{tabular}
  \caption{List of configs for the main training phase.}
  \label{tab:transformer}
\end{table}

\end{document}